\documentclass[pra,aps,twocolumn,floatfix,showpacs]{revtex4}
\usepackage{graphics}
\begin{document}
\title{Mixtures of ultracold fermions with unequal masses}
\author{M. Iskin and C. A. R. S{\'a} de Melo}
\affiliation{School of Physics, Georgia Institute of Technology, Atlanta, Georgia 30332, USA}
\date{\today}

\begin{abstract}
We analyze the phase diagram of superfluidity for
two-species fermion mixtures from the Bardeen-Cooper-Schrieffer (BCS) to 
Bose-Einstein condensation (BEC) limit as a function of scattering parameter,
population imbalance and mass anisotropy. 
We identify regions corresponding to normal, or uniform/non-uniform 
superfluid phases, and discuss topological quantum phase transitions 
in the BCS, unitarity and BEC limits.
We derive the Ginzburg-Landau equation near the critical temperature, 
and show that it describes a dilute mixture of paired and unpaired fermions in the BEC limit. 
We also obtain the zero temperature low frequency and long wavelength
collective excitation spectrum, and recover the Bogoliubov 
relation for weakly interacting dilute bosons in the BEC limit. 
Lastly, we discuss the effects of harmonic traps and the resulting density profiles
in the BEC regime.

\end{abstract}
\maketitle

\section{Introduction}
\label{sec:introduction}

The evolution from Bardeen-Cooper-Schrieffer (BCS) state to Bose-Einstein condensation (BEC)
is an important topic of current research for condensed matter, 
nuclear, atomic and molecular physics communities.
Recent advances in atomic physics have allowed for the study of superfluid properties 
in symmetric two-component fermion superfluids (equal mass and equal populations) as a function 
of scattering length,
where the theoretically predicted crossover from BCS to BEC was observed~\cite{leggett,nsr,carlos}.

Since the population of each component as well as the interaction strength between two components 
are experimentally tunable, these knobs enabled the study of the BCS to BEC evolution in 
asymmetric two-component fermion superfluids (equal mass but unequal populations)~\cite{mit,rice}.
In contrast with the crossover physics found in the symmetric case,
these experiments have demonstrated the existence of phase transitions between  
normal and superfluid phases, as well as phase separation between superfluid (paired)
and normal (excess) fermions as a function of population imbalance
~\cite{liu,yip-stability,bedaque,carlson,pao,son,sheehy,ho,bulgac,cheng,machida,pieri,torma,yi,chevy,silva,haque}.

Arguably one of the next frontiers of exploration in cold Fermi gases
is the study of asymmetric two-component fermion superfluidity 
(unequal masses and equal or unequal populations)
in two-species fermion mixtures from the BCS to the BEC limit.
Earlier works on two-species fermion mixtures were limited to the
BCS regime~\cite{liu,yip-stability,bedaque,lianyi}. 
However, very recently, the evolution from BCS to BEC was preliminarily addressed in homogenous systems 
as a function of population imbalance and scattering length 
especially for $^6$Li and $^{40}$K mixture and also for other mixtures including
$^6$Li and $^{87}$Sr and $^{40}$K and $^{87}$Sr mixtures~\cite{iskin-mixture, iskin-mixture2, parish}.
In addition, the superfluid phase diagram of trapped systems 
at unitarity was also analyzed as a function of population imbalance 
and mass anisotropy~\cite{pao-mixture}.

In this manuscript, we study the BCS to BEC evolution of 
asymmetric two-component fermion superfluids
as a function of population imbalance and mass anisotropy, and extend our previous
works~\cite{iskin-mixture, iskin-mixture2} in this area.
Our main results are as follows.
For a homogeneous system, we analyze the zero temperature phase diagram for ultra cold fermion mixtures
as a function of mass anisotropy and population imbalance.
We identify regions corresponding to normal, uniform or non-uniform 
superfluid phases, and discuss topological quantum phase transitions 
in the BCS, unitarity and BEC limits.
We derive the Ginzburg-Landau theory near the critical temperature, 
and show that it describes a dilute mixture of weakly interacting paired and unpaired fermions in the BEC limit. 
We also derive the zero temperature low frequency and long wavelength
collective excitation spectrum for zero population imbalance, and recover the Bogoliubov 
relation for weakly interacting dilute bosons in the BEC limit. In addition, we describe analytically 
the phase separation boundaries of the resulting Bose-Fermi mixture 
of paired fermions and unpaired fermions in the BEC limit.
Lastly, we discuss the effects of harmonic traps and the resulting density profiles
of paired and unpaired fermions in the BEC regime.

The rest of the paper is organized as follows. In Sec.~\ref{sec:fif}, 
we introduce the imaginary-time functional integration 
formalism, and obtain the self-consistency (order parameter and number) equations. 
In Sec.~\ref{sec:saddle}, we discuss the evolution from
BCS to BEC superfluidity at zero temperature within the saddle point approximation,
and we analyze the order parameter, chemical potential, and stability
of the saddle-point solutions.
In Sec.~\ref{sec:fluctuations}, we discuss gaussian fluctuations near the critical
temperature, and we obtain the low energy collective excitations and  
corrections to the saddle point phase diagram in the BEC region at zero temperature. In addition, we discuss
the effects of harmonic traps on the density profile of paired and unpaired fermions 
at zero temperature.
A summary of our conclusions is given in Sec.~\ref{sec:conclusions}.
Lastly, we present in Appendices~\ref{sec:app.a},~\ref{sec:app.b}, and~\ref{sec:app.c}
the elements of the inverse fluctuation matrix, and 
their low frequency and long wavelength expansion coefficients 
at finite and zero temperatures.

\section{Functional Integral Formalism}
\label{sec:fif}

To describe a dilute two-species Fermi gas in three dimensions, 
we start from the Hamiltonian ($\hbar = 1$)
\begin{equation}
\label{eqn:hamiltonian}
H = \sum_{\mathbf{k},\sigma} \xi_{\mathbf{k},\sigma} a_{\mathbf{k},\sigma}^\dagger a_{\mathbf{k},\sigma} + 
\sum_{\mathbf{k},\mathbf{k'},\mathbf{q}} U (\mathbf{k},\mathbf{k'})
b_{\mathbf{k},\mathbf{q}}^\dagger b_{\mathbf{k'},\mathbf{q}}, 
\end{equation}
where the pseudo-spin $\sigma$ labels both the type and hyperfine states of atoms
represented by the creation operator $ a_{\mathbf{k},\sigma}^\dagger$, and
$b_{\mathbf{k},\mathbf{q}}^\dagger = a_{\mathbf{k}+\mathbf{q}/2,\uparrow}^\dagger 
a_{-\mathbf{k}+\mathbf{q}/2,\downarrow}^\dagger$.
Here, $\xi_{\mathbf{k},\sigma}= \epsilon_{\mathbf{k},\sigma} - \mu_\sigma$,
where $\epsilon_{\mathbf{k},\sigma} = k^2/(2m_\sigma)$ is the energy
and $\mu_\sigma$ is the chemical potential of the fermions.
Notice that we allow for the fermions to have different masses $m_{\sigma}$ and
different populations controlled by independent chemical potentials $\mu_{\sigma}$.
The attractive fermion-fermion interaction $U (\mathbf{k},\mathbf{k'})$ 
can be written in a separable form as
$
U (\mathbf{k},\mathbf{k'}) =  - g \Gamma^*_\mathbf{k} \Gamma_\mathbf{k'} 
$
where $g > 0$, and $\Gamma_\mathbf{k} = 1$ for the zero-range s-wave contact interaction
considered in this manuscript.

\subsection{Effective action}
\label{sec:effectiveaction}

The gaussian effective action for $H$ is
\begin{equation}
\label{eqn:eff-action}
S_{\rm Gauss} = S_0 + \frac{\beta}{2} \sum_{q} \bar{\Lambda}^\dagger(q) \mathbf{F}^{-1}(q) \bar{\Lambda}(q),
\end{equation}
where $\beta = 1/T$ and $q=(\mathbf{q},v_\ell)$ with bosonic Matsubara frequency $v_\ell=2\ell\pi/\beta$.
Here, the saddle point action is given by
\begin{equation}
S_0 = \beta \frac{|\Delta_0|^2}{g} - \sum_{p} \rm{Tr} \ln [\mathbf{G}^{\rm sp}(p)/\beta]^{-1},
\end{equation}
where $p=(\mathbf{k},w_\ell)$ with fermionic Matsubara frequency $w_\ell=(2\ell+1)\pi/\beta$,
$
\Delta_\mathbf{k} = \Delta_0\Gamma_\mathbf{k}
$
is the order parameter, and
$(\mathbf{G}^{\rm sp})^{-1} (p)$ is the inverse fermion propagator.
The matrix $(\mathbf{G}^{\rm sp})^{-1} (p)$ is defined by
\begin{equation}
(\mathbf{G}^{\rm sp})^{-1} (p) = \left( \begin{array}{cc} 
iw_\ell - \xi_{\mathbf{k},\uparrow} & \Delta_{\mathbf{k}} \\
\Delta_{\mathbf{k}}^* & iw_\ell + \xi_{\mathbf{k},\downarrow}
\end{array}\right).
\end{equation}
Furthermore, the vector $\bar{\Lambda}^\dagger(q)$ is the order parameter fluctuation field
and $\mathbf{F}^{-1}(q)$ is the inverse fluctuation propagator. 
The matrix elements of $\mathbf{F}^{-1}(q)$ are given by
\begin{eqnarray}
\label{eqn:f1-matrix}
\mathbf{F}^{-1}_{1,1} (q) 
&=& \frac{V}{g} - \frac{1}{\beta} \sum_p \mathbf{G}^{\rm sp}_{\uparrow,\uparrow}(\frac{q}{2} + p) 
\mathbf{G}^{\rm sp}_{\downarrow,\downarrow}(\frac{q}{2} - p) |\Gamma_\mathbf{k}|^2,
\label{eqn:fluct.F11}
\\
\label{eqn:f2-matrix}
\mathbf{F}^{-1}_{1,2} (q)
&=& 
\frac{1}{\beta} \sum_p \mathbf{G}^{\rm sp}_{\uparrow,\downarrow} (\frac{q}{2} + p) 
\mathbf{G}^{\rm sp}_{\uparrow,\downarrow}(\frac{q}{2} - p) |\Gamma_\mathbf{k}|^2.
\label{eqn:fluct.F12}
\end{eqnarray}
Notice that $\mathbf{F}^{-1}_{2,1}(q) = (\mathbf{F}^{-1}_{1,2})^*(q)$ and
$\mathbf{F}^{-1}_{2,2}(q) = \mathbf{F}^{-1}_{1,1}(-q)$.
These matrix elements are described further in appendix~\ref{sec:app.a}.
The fluctuation term in the action leads to a correction
to the thermodynamic potential, which can be written as 
$\Omega_{{\rm Gauss}} = \Omega_0 + \Omega_{{\rm fluct}}$ with 
$\Omega_0 = S_0/\beta$ and
$\Omega_{{\rm fluct}} = (1/\beta)\sum_{q}\ln\det[\mathbf{F}^{-1}(q)/\beta]$.
Next, we derive the self-consistency equations.

\subsection{Self-consistency equations}
\label{sec:self-consistency}

The saddle point condition $\delta S_0 /\delta \Delta_0^* = 0$ or the relation 
\begin{equation}
\Delta_{\mathbf{k'}} = - \frac{1}{\beta} \lim_{\tau \to 0}
\sum_p U (\mathbf{k},\mathbf{k'}) \mathbf{G}^{\rm sp}_{\uparrow,\downarrow} (p) \exp(i w_\ell \tau)
\end{equation}
leads to an equation for the order parameter
\begin{equation}
\label{eqn:order-parameter}
\frac{1}{g} = \sum_{\mathbf{k}} \frac{|\Gamma_\mathbf{k}|^2} {2E_{\mathbf{k},+}} {\cal X}_{\mathbf{k},+},
\end{equation}
where
$
{\cal X}_{\mathbf{k},\pm} = ( {\cal X}_{\mathbf{k},1} \pm {\cal X}_{\mathbf{k},2} ) / 2
$
with
$
{\cal X}_{\mathbf{k},s} = \tanh(\beta E_{\mathbf{k},s}/2).
$
Notice that, at low temperatures $T \approx 0$,
$\theta(-E_{\mathbf{k,s}}) = \lim_{\beta \to \infty} {\cal X}_{\mathbf{k},s}$,
where $\theta(x)$ is the Heaviside function.
Here,
$
E_{\mathbf{k},\pm} = (E_{\mathbf{k},1} \pm E_{\mathbf{k},2})/2
$
and
$
\xi_{\mathbf{k},\pm} = (\xi_{\mathbf{k},\uparrow} \pm \xi_{\mathbf{k},\downarrow})/2
= k^2/(2m_\pm) - \mu_\pm,
$
where
\begin{eqnarray}
\label{eqn:quasiparticle-energies}
E_{\mathbf{k},1} &=& (\xi_{\mathbf{k},+}^2 + |\Delta_\mathbf{k}|^2)^{1/2} + \xi_{\mathbf{k},-}, \\
E_{\mathbf{k},2} &=& (\xi_{\mathbf{k},+}^2 + |\Delta_\mathbf{k}|^2)^{1/2} - \xi_{\mathbf{k},-}
\end{eqnarray}
are the quasiparticle and negative of the quasihole energies respectively,
$
m_\pm = 2m_\uparrow m_\downarrow/(m_\downarrow \pm m_\uparrow)
$
and 
$
\mu_\pm = (\mu_\uparrow \pm \mu_\downarrow)/2.
$
Notice that $m_+$ is twice the reduced mass of the $\uparrow$ 
and $\downarrow$ fermions, and that the equal mass case corresponds 
to $|m_-| \to \infty$. 
As usual, we eliminate $g$ in favor of the scattering length $a_F$ via the relation
\begin{equation}
\label{eqn:interaction-scattering}
\frac{1}{g} = - \frac{m_+ V} {4\pi a_F} + \sum_{\mathbf{k}} \frac{|\Gamma_\mathbf{k}|^2} {2\epsilon_{\mathbf{k},+}},
\end{equation}
where 
$
\epsilon_{\mathbf{k},\pm} = (\epsilon_{\mathbf{k},\uparrow} \pm \epsilon_{\mathbf{k},\downarrow})/2.
$

The order parameter equation has to be solved self-consistently with number equations
$N_\sigma = -\partial \Omega/\partial {\mu_\sigma}$ which have two contributions 
$N_\sigma = N_{0,\sigma} + N_{{\rm fluct},\sigma}$.
The first term $N_{0,\sigma} = - \partial \Omega_0/\partial {\mu_\sigma}$ or the relation
\begin{equation}
N_{0,\sigma} = \frac{\gamma_\sigma}{\beta} \lim_{\tau \to 0} \sum_p 
\mathbf{G}^{\rm sp}_{\sigma,\sigma} (p) \exp(i \gamma_\sigma w_\ell \tau) 
\end{equation}
leads to the saddle point contribution, and is given by
\begin{equation}
\label{eqn:numbereqn}
N_{0,\sigma} = \sum_{\mathbf{k}} 
\left( \frac{1 - \gamma_\sigma {\cal X}_{\mathbf{k},-}} {2}
- \frac{\xi_{\mathbf{k},+}}{2E_{\mathbf{k},+}} {\cal X}_{\mathbf{k},+}
\right).
\end{equation}
Here, $\gamma_\uparrow = + 1$ and $\gamma_\downarrow = -1$.
Similarly, the second term $N_{{\rm fluct},\sigma} = -\partial \Omega_{{\rm fluct}}/\partial {\mu_\sigma}$
leads to the fluctuation contribution, and is given by
\begin{equation}
N_{{\rm fluct},\sigma} = - \frac{1}{\beta} \sum_{q}
\frac{\partial [\det \mathbf{F}^{-1}(q)] / \partial {\mu_\sigma}} 
{\det \mathbf{F}^{-1}(q)}.
\end{equation}
While the saddle point contribution is sufficient for a semi-quantitative 
analysis at zero temperature ($T \approx 0$),
inclusion of the fluctuation contribution is necessary to recover BEC physics
at finite temperatures ($T \to T_c$).

When populations of the pseudo-spin components are balanced 
($N_\uparrow = N_\downarrow$), the results for $|\Delta_0|$ and $\mu_+$
($\mu_-$ is irrelevant in this case) of $m_\uparrow \ne m_\downarrow$ case
can be obtained from the results for $|\Delta_0|$ and $\mu$ of 
$m = m_\uparrow = m_\downarrow$ case via a substitution of $m \to m_+$.
However, when populations of the pseudo-spin components are imbalanced 
($N_\uparrow \ne N_\downarrow$), we need to solve all
three self-consistency equations, since population imbalance is achieved
when either $E_{\mathbf{k},1}$ or $E_{\mathbf{k},2}$ is negative 
in some regions of momentum space, which is discussed next.

\section{Saddle point results}
\label{sec:saddle}

At low temperatures $T \approx 0$, the saddle point self-consistency 
(order parameter and number) equations are sufficient to describe 
the evolution of superfluidity from the BCS to the BEC limit.
In this section, we analyze amplitude of the order parameter
$|\Delta_0|$ and chemical potentials $\mu_\sigma$ as a function of 
mass anisotropy $m_r = m_\uparrow / m_\downarrow$ and population 
imbalance $P = N_-/N_+$ for a set of fixed scattering parameters $1/(k_{F,+} a_F)$.
Here $N_\pm = (N_\uparrow \pm N_\downarrow)/2$ and 
$k_{F,\pm}^3 = (k_{F,\uparrow}^3 \pm k_{F,\downarrow}^3)/2$. Because of the
parabolic dispersion relation, the density of $\uparrow$ fermions is 
$n_{\uparrow} = k_{F,{\uparrow}}^3/(6\pi^2)$
and the density of $\downarrow$ fermions is 
$N_{\downarrow} = k_{F,{\downarrow}}^3/(6\pi^2)$.
Here, the Fermi momenta $k_{F,{\uparrow}}$ and $k_{F,{\downarrow}}$ are determined from the Fermi energies
$\epsilon_{F,\sigma} = k_{F,\sigma}^2/(2 m_{\sigma})$. Therefore, the total
fermion density is 
\begin{equation}
\label{eqn:n-total}
n = n_{\uparrow} + n_{\downarrow} = \frac{k_{F,+}^3}{3\pi^2}. 
\end{equation}

\begin{figure} [htb]
\centerline{\scalebox{0.6}{\includegraphics{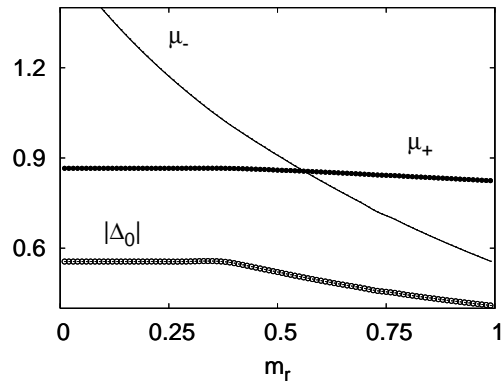} }}
\caption{\label{fig:gap} 
Plots of $|\Delta_0|$, $\mu_+$ and $\mu_-$ (in units of $\epsilon_{F,+}$)
as a function of $m_r$, when $P = 0.5$ and $1/(k_{F,+} a_F) = 0$ (unitarity limit).
Notice the presence of small cusps in $|\Delta_0|$ and $\mu_+$ when $m_r \approx 0.4$ 
which signal a topological quantum phase transition discussed below.
These cusps are more pronounced for higher $|P|$ (not shown), but are best seen in 
this figure at small grazing angles.
}
\end{figure}

Using the notations described in the preceding paragraph, we can 
solve the self-consistency Eqs.~(\ref{eqn:order-parameter}),~(\ref{eqn:interaction-scattering})
and~(\ref{eqn:numbereqn}). For instance, in  Fig.~\ref{fig:gap}, we show self-consistent solutions of
$|\Delta_0|$, $\mu_+$ and $\mu_-$ (in units of $\epsilon_{F,+}$) 
as a function of $m_r$ when $P = 0.5$ and $1/(k_{F,+} a_F) = 0$.
Here, $\epsilon_{\rm F,\pm} = k_{F,\pm}^2/(2m_{\pm})$.
With increasing mass anisotropy (or decreasing $m_r$), we find that both $|\Delta_0|$ 
and $\mu_+$ increase until $m_r \approx 0.4$. 
However, further decrease in $m_r$ beyond $m_r \approx 0.4$ leads to a 
saturation of both $|\Delta_0|$ and $\mu_+$ with a small cusp in both quantities.
The cusp is best seen in Fig.~\ref{fig:gap} at small grazing angles.
Therefore, the evolution from $m_r = 1$ to $m_r \to 0$ is non-analytic 
when $m_r \approx 0.4$, and the evolution is not a crossover.
These cusps in $|\Delta_0|$ and $\mu_+$ are more pronounced for 
higher $|P|$, and they signal a topological quantum phase transition discussed below.
Notice that, for $P = 0$, the evolution of $|\Delta_0|$ and $\mu_+$ 
is analytic for all $m_r$, and the evolution is just a crossover. 

Next, we discuss the stability of uniform superfluidity using two criteria
~\cite{pao, gubankova, lianyi2, sheehy2, pao-stability}:
positive definite compressibility $\mathbf{\kappa}(T)$ 
and superfluid density $\mathbf{\rho}(T)$ matrices.

\subsection{Stability of uniform superfluidity}
\label{sec:stability}

In order to analyze the phase diagram at $T = 0$, 
we solve the saddle point self-consistency (order parameter and number) 
equations for all $P$ and $m_r$ for a set of $1/(k_{F,+} a_F)$, 
and check the stability of saddle point solutions for the
uniform superfluid phase using two criteria.

The first criterion requires that the compressibility matrix 
$\mathbf{\kappa}(T)$ is positive definite,
where the elements of $\mathbf{\kappa}(T)$  are 
$
\kappa_{\sigma,\sigma'} (T) = - \partial^2 \Omega_0 / (\partial \mu_\sigma \partial \mu_{\sigma'}).
$
This criterion is related (identical) to the condition that the curvature
\begin{equation}
\frac{\partial^2 \Omega_0} {\partial \Delta_0^2} = 
\sum_{\mathbf{k}} |\Delta_{\mathbf{k}}|^2 
\left( 
\frac{{\cal X}_{\mathbf{k}, +}}{E_{\mathbf{k},+}^3} -
\beta \frac{{\cal Y}_{\mathbf{k}, +}}{2E_{\mathbf{k},+}^2}  
\right).
\label{eqn:curvature}
\end{equation}
of the saddle point thermodynamic potential $\Omega_0$ with respect to the saddle point parameter $\Delta_0$ 
needs to be positive~\cite{lianyi2, pao-stability}. 
Here,
$
{\cal Y}_{\mathbf{k},\pm} = ({\cal Y}_{\mathbf{k},\uparrow} \pm  {\cal Y}_{\mathbf{k},\downarrow})/2
$
with ${\cal Y}_{\mathbf{k},s} = {\rm sech}^2 (\beta E_{ \mathbf {k}, s}/2)$. 
Notice that, at low temperatures $T \approx 0$, 
$\delta(E_{\mathbf{k,s}}) = \lim_{\beta \to \infty} \beta {\cal Y}_{\mathbf{k},s}/4$
where $\delta(x)$ is the delta function.
When at least one of the eigenvalues of $\mathbf{\kappa} (T)$, 
or the curvature $\partial^2 \Omega_0 /\partial \Delta_0^2$ is
negative, the uniform saddle point solution does not correspond to a minimum of $\Omega_0$, 
and a non-uniform superfluid phase is favored.

The second criterion requires that superfluid density matrix $\mathbf{\rho}(T)$ 
to be positive definite, where the elements of $\mathbf{\rho}(T)$ are defined by
$
\rho_{i,j}(T) = (1/\beta) \sum_p \lbrace {\rm Tr}[\mathbf{G}^{\rm sp}(p) \widetilde{\mathbf{M}}] \delta_{i,j}
+ k_i k_j {\rm Tr}[\mathbf{G}^{\rm sp}(p) \mathbf{G}^{\rm sp}(p)] \rbrace,
$
where $\widetilde{\mathbf{M}}$ is a diagonal mass matrix with 
elements $\widetilde{\mathbf{M}}_{i,j} = \gamma_\sigma m_\sigma \delta_{i,j}$
with $\gamma_\uparrow = 1$ and $\gamma_\downarrow = -1$, and $\delta_{i,j}$ is the Kronecker delta.
After the evaluation of the fermionic Matsubara frequency, we obtain
\begin{equation}
\rho_{ij}(T) = (m_\uparrow N_\uparrow + m_\downarrow N_\downarrow)\delta_{i,j}
- \frac{\beta}{2} \sum_{\mathbf{k}} k_i k_j {\cal Y}_{\mathbf{k},+}.
\label{eqn:sd}
\end{equation}
When at least one of the eigenvalues of $\mathbf{\rho}(T)$ is negative, a spontaneously generated gradient
of the phase of the order parameter appears, leading to a non-uniform superfluid phase.
Notice that the $\mathbf{\rho} (T)$ matrix is reducible to the scalar
\begin{equation}
\rho_0(T) = m_\uparrow N_\uparrow + m_\downarrow N_\downarrow
- \frac{\beta}{6} \sum_{\mathbf{k}} k^2 {\cal Y}_{\mathbf{k},+},
\end{equation}
in the s-wave case.

Before discussing ground state phase diagrams, we would like to add an additional criterion to fine
tune the classification of the various phases that emerge as a result of unequal masses, interactions,
and population imbalance. For this purpose we discuss next, the quasiparticle excitation spectrum
and its connection to topological quantum phase transitions.

\subsection{Topological quantum phase transitions}
\label{sec:qpt}

\begin{figure} [htb]
\centerline{\scalebox{0.325}{\includegraphics{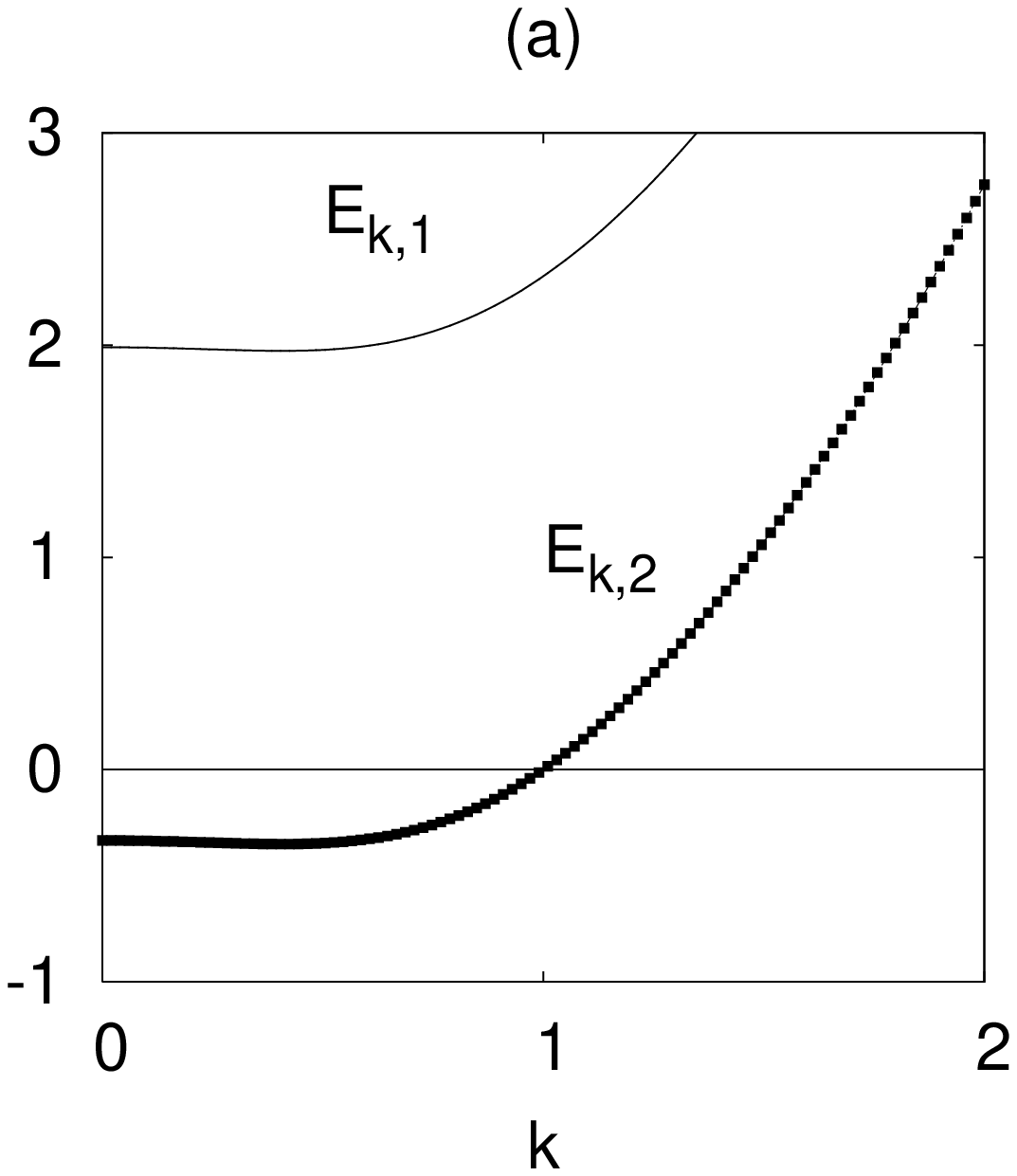} \includegraphics{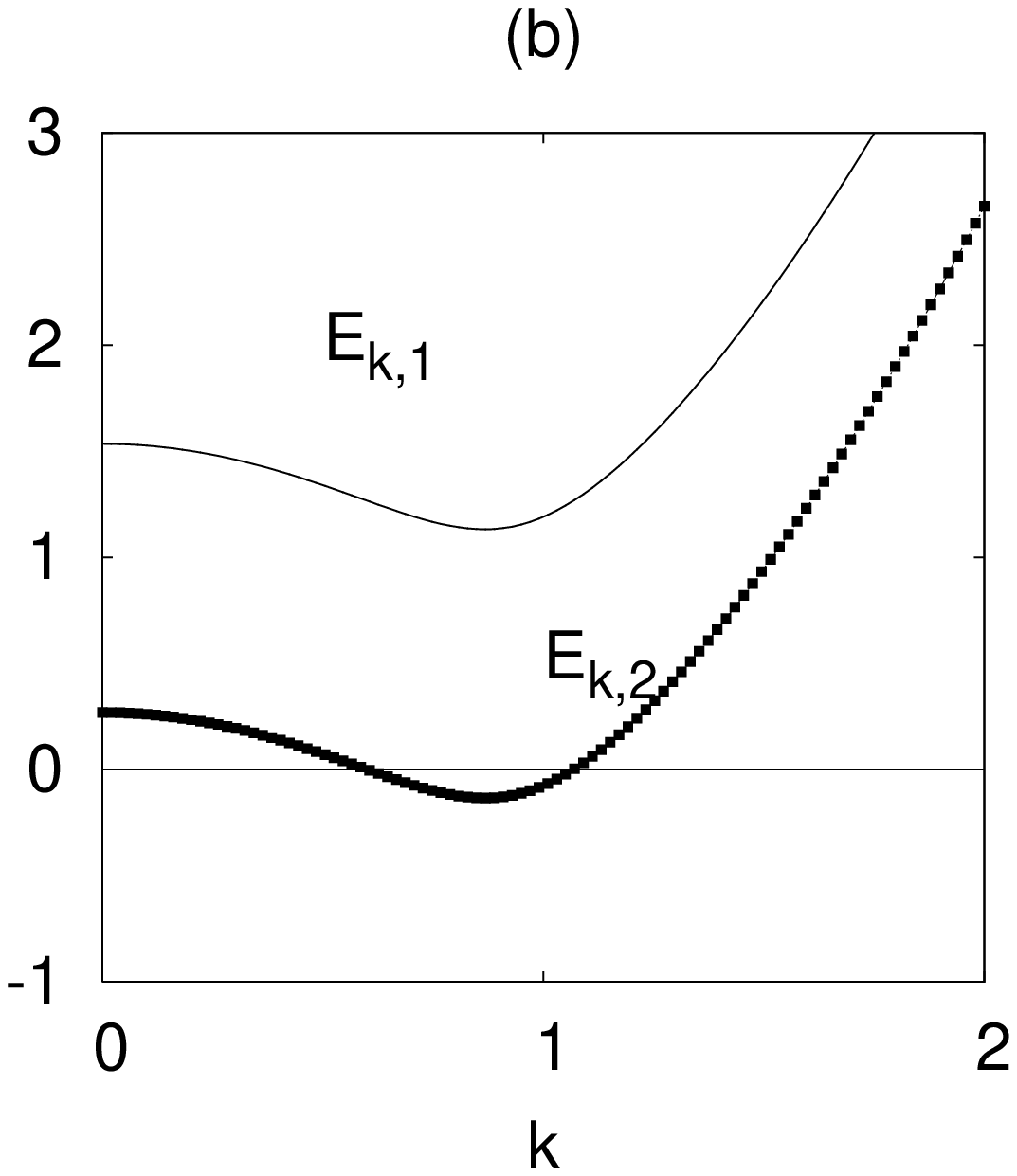}}}
\centerline{\scalebox{0.33}{\includegraphics{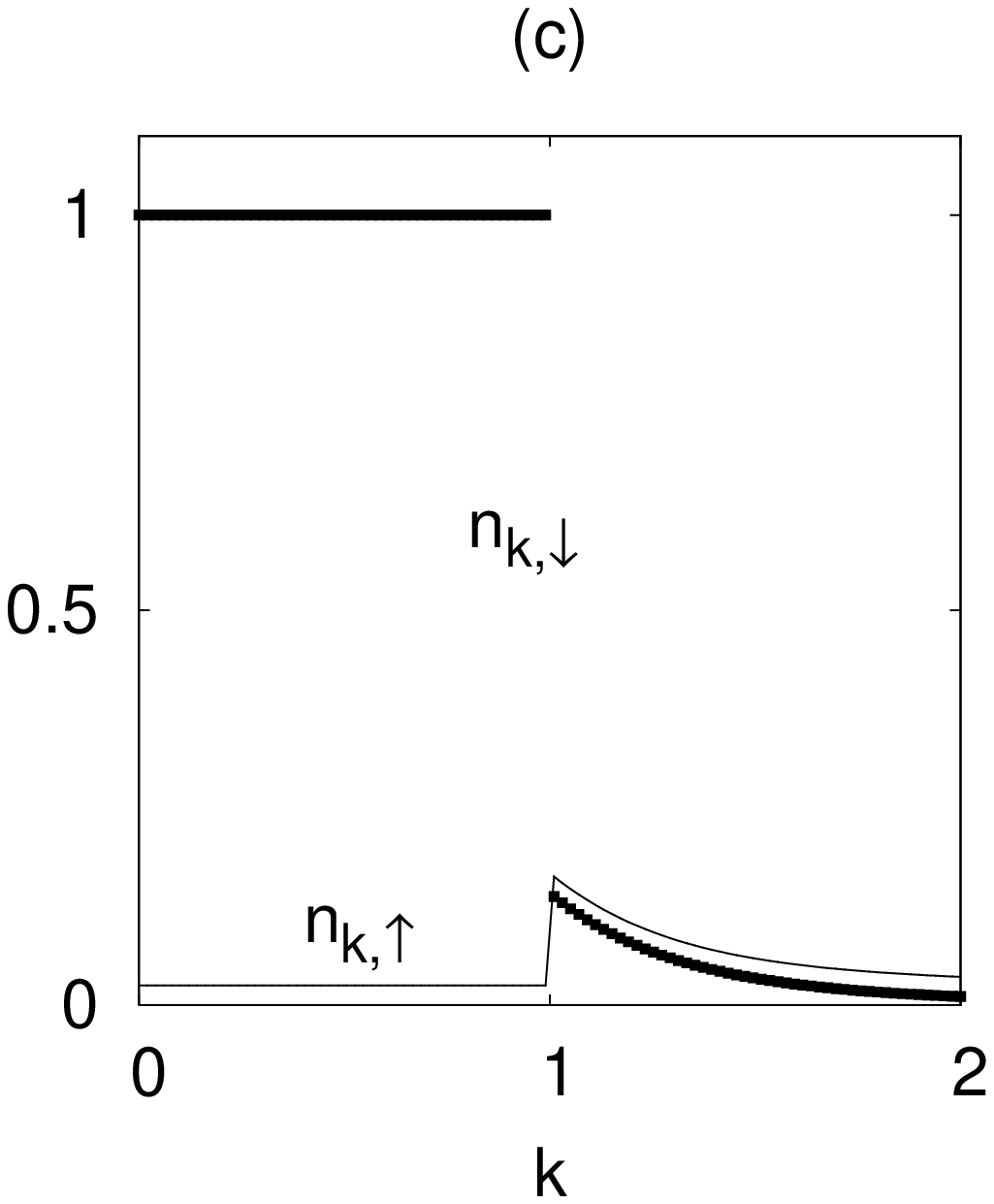} \includegraphics{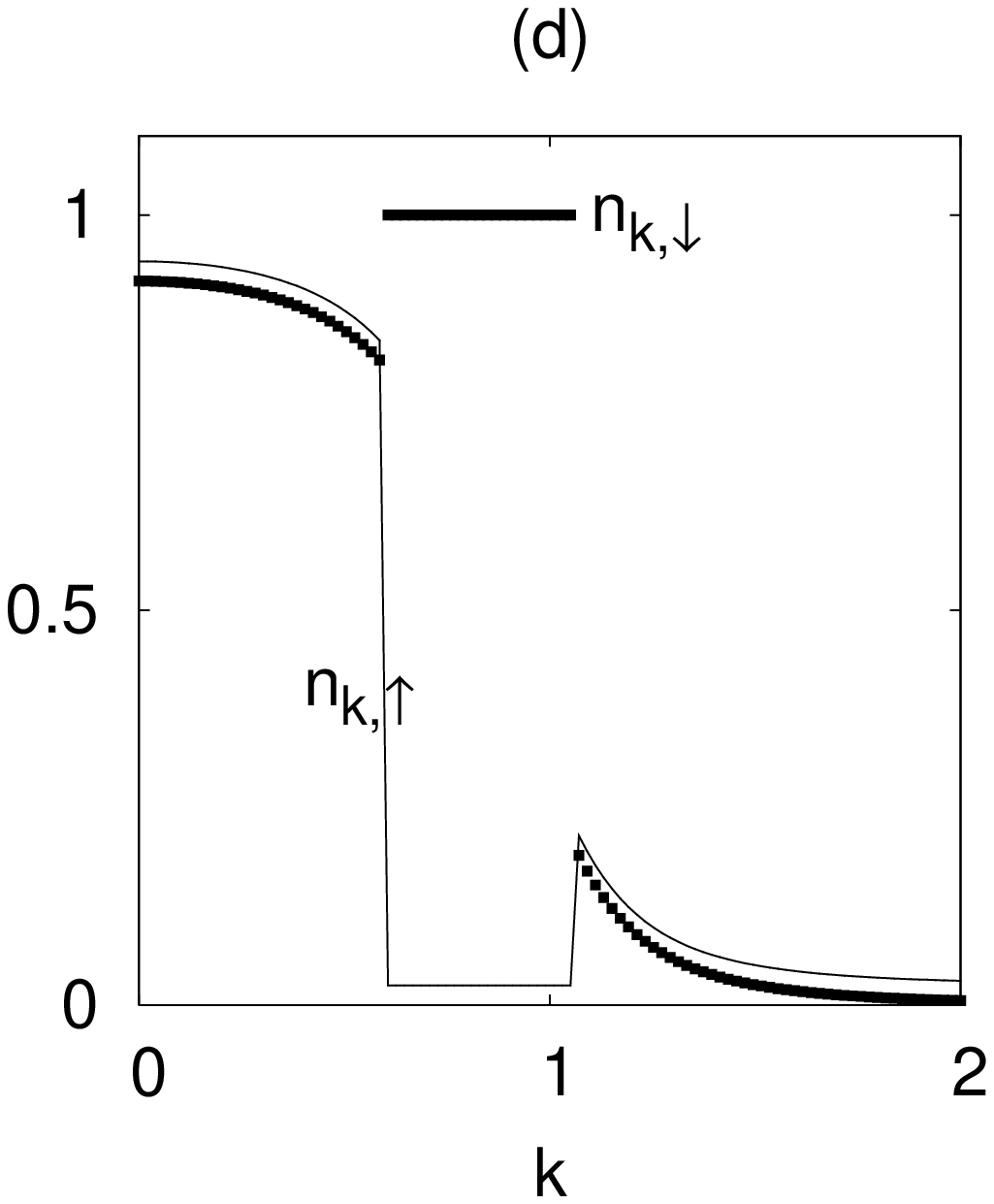}}}
\caption{\label{fig:qpt}
Schematic plots of $E_{\mathbf{k},1}$ (thin lines) and $E_{\mathbf{k},2}$ (thick lines) versus $k$
for (a) phase I and (b) phase II. Corresponding momentum distributions 
$n_{\mathbf{k},\uparrow}$ (thin lines) and $n_{\mathbf{k},\downarrow}$ (thick lines) are 
shown in (c) phase I and (d) phase II. Notice that the zero of $n_{\mathbf{k},\uparrow}$ is slightly 
shifted upwards from the zero of $n_{\mathbf{k},\downarrow}$, for better visualization.
}
\end{figure}

The excitation spectrum of quasiparticles is determined by
energies $E_{\mathbf{k},1}$ and $E_{\mathbf{k},2}$ defined in 
Eq.~(\ref{eqn:quasiparticle-energies}). Using these relations,
one can identify surfaces in momentum space where these energies
have zeros, indicating that the quasiparticle excitation spectrum 
changes from a gapped to a gapless phase, with a corresponding change in the 
momentum distribution as well. These changes in the Fermi surfaces of
quasiparticles are topological in nature. Thus, we identify topological quantum phase transitions associated 
with the disappearance or appearance of momentum space regions
of zero quasiparticle energies when either $1/(k_{F,+} a_F)$, $P$, 
and/or $m_r$ is changed. These topological transitions are shown as dotted lines
in Figs.~\ref{fig:phase.bcs} through \ref{fig:phase.bec}.
These phases are characterized by the number of zeros 
of $E_{\mathbf{k},1}$ and $E_{\mathbf{k},2}$
(zero energy surfaces in momentum space) such that
I) $E_{\mathbf{k},1}$ has no zeros and $E_{\mathbf{k},2}$ has only one, and
II) $E_{\mathbf{k},1}$ has no zeros and $E_{\mathbf{k},2}$ has two zeros.
In the case of unequal masses, the superfluid has gapless quasiparticle excitations in both phases I and II, 
when there is population imbalance, e.g., $N_\uparrow \ne N_\downarrow$.
We illustrate these cases in Figs.~\ref{fig:qpt}(a) and~\ref{fig:qpt}(b), respectively, for $N_\uparrow > N_\downarrow$.
The case of no population imbalance $P  = 0$ ($N_\uparrow =  N_\downarrow$) corresponds to case III, 
where $E_{\mathbf{k},1}$ and $E_{\mathbf{k},2}$ have no zeros and are always positive, and thus the 
superfluid has always gapped quasiparticle excitations.

The transitions among phases I, II and III indicate a change in 
topology in the lowest quasiparticle band, similar to the Lifshitz transition in 
ordinary metals~\cite{lifshitz-60} and non-swave superfluids~\cite{volovik-92,borkowski-98,duncan-00,botelho-05,iskinprl}. 
The topological transition here is unique, because it involves an s-wave superfluid, 
and could be potentially observed for the first time through the measurement of the momentum distribution 
or thermodynamic properties. Notice that the topological transition occurs without changing
the symmetry of the order parameter as the Landau classification demands for ordinary phase transitions.
However, thermodynamic signatures of the topological transition are present at low temperatures in 
the compressibility, specific heat, superfluid density, etc..., because the quasiparticle excitation 
spectrum changes dramatically. The temperature dependence of the quasiparticle contributions to these properties
are exponentially activated [$\sim \exp(-E_g/T)$] for the gapped phase (III), or have a power
law behavior (with different powers of $T$) in the gapless phases (I and II).

The zero temperature momentum distributions 
\begin{equation}
\label{eqn:momentum-distribution}
n_{\mathbf{k},\sigma} = \frac{1 - \gamma_\sigma {\cal X}_{\mathbf{k},-}} {2}
- \frac{\xi_{\mathbf{k},+}}{2E_{\mathbf{k},+}} {\cal X}_{\mathbf{k},+}
\end{equation}
for phases I and II are extracted from Eq.~(\ref{eqn:numbereqn}).
For momentum space regions where $E_{{\bf k},1} > 0$ and $E_{{\bf k},2} > 0$,
the corresponding momentum distributions are equal $n_{{\bf k},\uparrow} =  n_{{\bf k},\downarrow}$.
However, when $E_{{\bf k},1} > 0$ and $E_{{\bf k},2} < 0$, then
$n_{{\bf k},\uparrow} = 0$ and $n_{{\bf k},\downarrow} = 1$. 
We illustrate these cases in Figs.~\ref{fig:qpt}(c) and~\ref{fig:qpt}(d) 
for parameters of Figs.~\ref{fig:qpt}(a) and~\ref{fig:qpt}(b), respectively.
Notice that the zero of $n_{\mathbf{k},\uparrow}$ is shifted slightly
upwards to distinguish it from $n_{\mathbf{k},\downarrow}$
in the regions of momentum space where $n_{\mathbf{k},\uparrow} = n_{\mathbf{k},\downarrow}$.
Although this topological transition is quantum ($T=0$) in nature, 
signatures of the transition should still be observed
at finite temperatures within the quantum critical region, where the momentum distributions
are smeared out due to thermal effects. Although the primary signature of this topological 
transition is seen in the momentum distribution, the isentropic $\kappa_S$ or isothermal $\kappa_T$ compressibilities 
and the speed of sound $c_s$ would have a cusp at the topological transition line 
similar to that encountered in $|\Delta_0|$ (see Fig.~\ref{fig:gap})
as a function of the mass anisotropy $m_r$. The cusp (discontinuous change in
slope) in $\kappa_S$, $\kappa_T$ or $c_s$ gets larger with increasing population imbalance. 

Having discussed the finer topological classification of possible superfluid phases, we take all the
criteria together (positive compressibility, positive curvature of thermodynamic potential, positive superfluid density,
and topological character) to present next the resulting ground state phase diagrams.

\subsection{Ground state phase diagrams}
\label{sec:phase}

\begin{figure} [htb]
\centerline{\scalebox{0.6}{\includegraphics{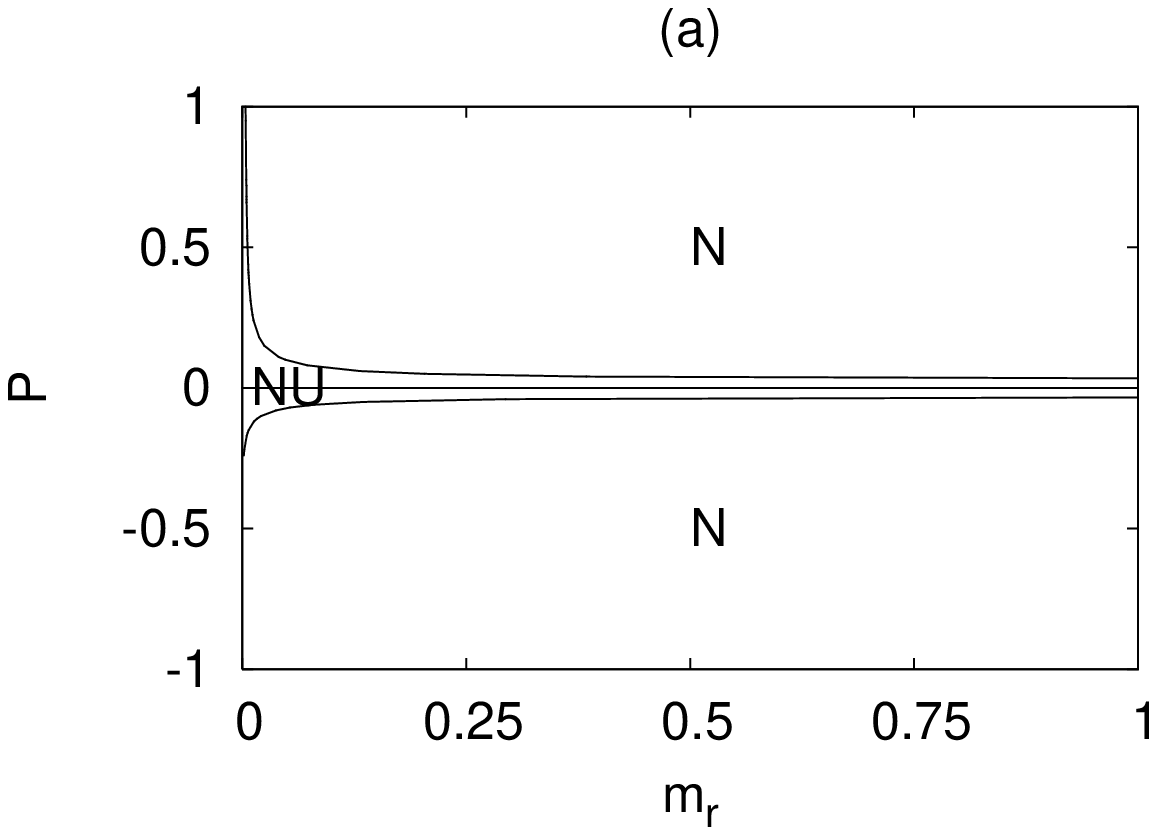} }}
\centerline{\scalebox{0.6}{\includegraphics{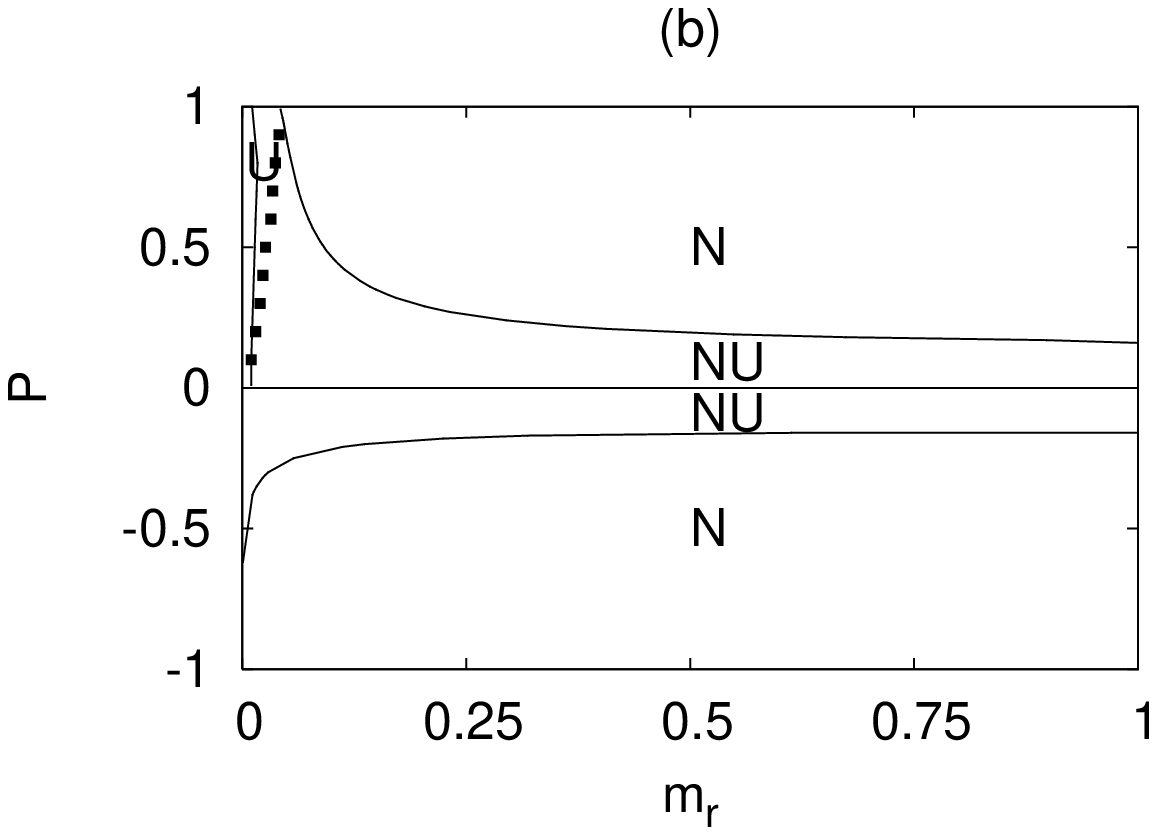} }}
\centerline{\scalebox{0.6}{\includegraphics{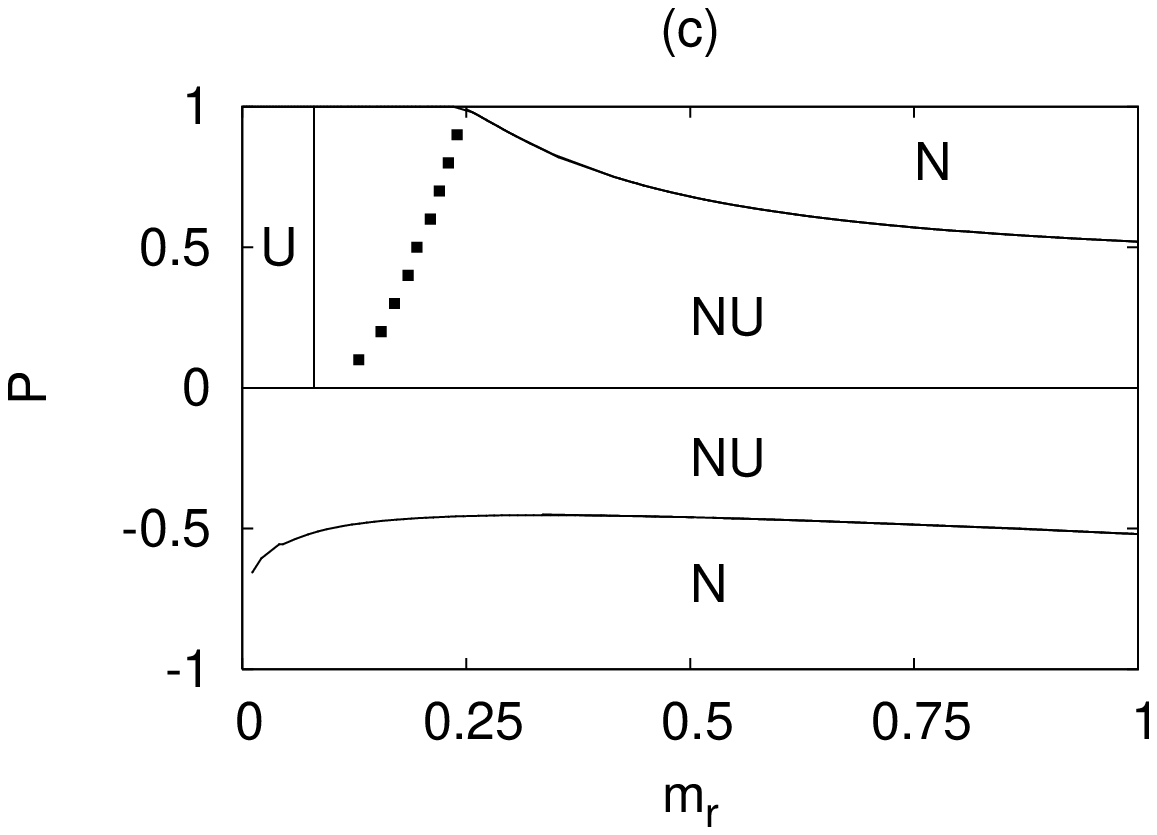} }}
\caption{\label{fig:phase.bcs} 
Phase diagram of $P = (N_\uparrow - N_\downarrow)/(N_\uparrow + N_\downarrow)$ 
versus $m_r = m_\uparrow / m_\downarrow$ on the BCS side 
when a) $1/(k_{F,+} a_F) = -2$ b) $1/(k_{F,+} a_F) = -1$ and c) $1/(k_{F,+} a_F) = -0.25$.
We show normal (N), uniform (U) or non-uniform (NU) superfluid phases.
The dotted line (black squares) separate topologically distinct regions.
}
\end{figure}

\begin{figure} [htb]
\centerline{\scalebox{0.6}{\includegraphics{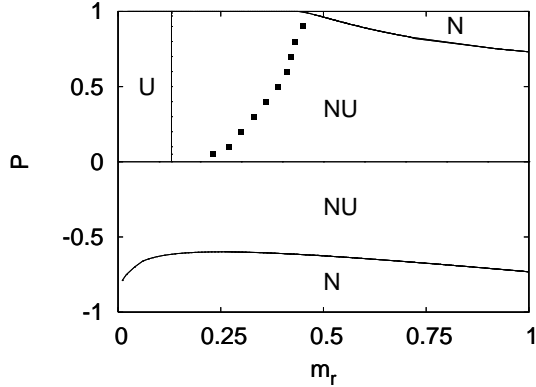} }}
\caption{\label{fig:phase.unitarity} 
Phase diagram of $P = (N_\uparrow - N_\downarrow)/(N_\uparrow + N_\downarrow)$ 
versus $m_r = m_\uparrow / m_\downarrow$ at the unitarity limit
when $1/(k_{F,+} a_F) = 0$.
We show normal (N), uniform (U) or non-uniform (NU) superfluid phases.
The dotted line (black squares) separate topologically distinct regions.
}
\end{figure}

\begin{figure} [htb]
\centerline{\scalebox{0.6}{\includegraphics{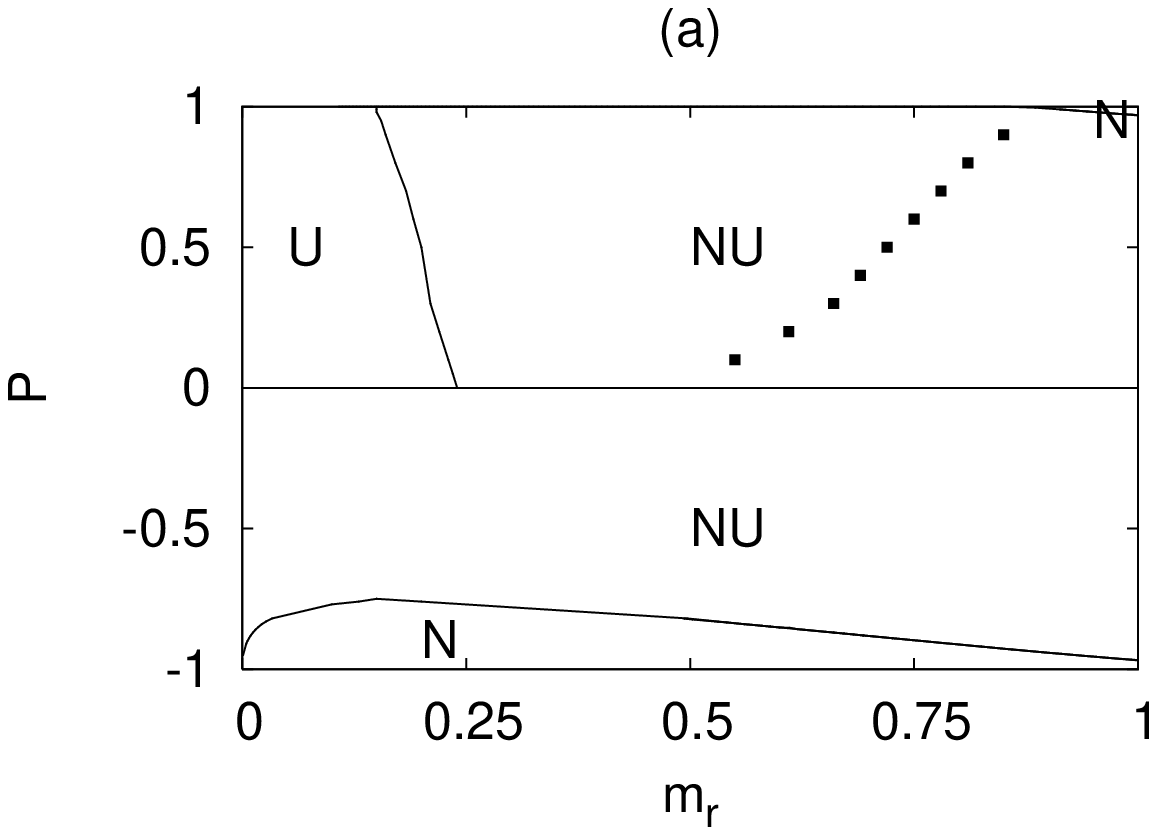} }}
\centerline{\scalebox{0.6}{\includegraphics{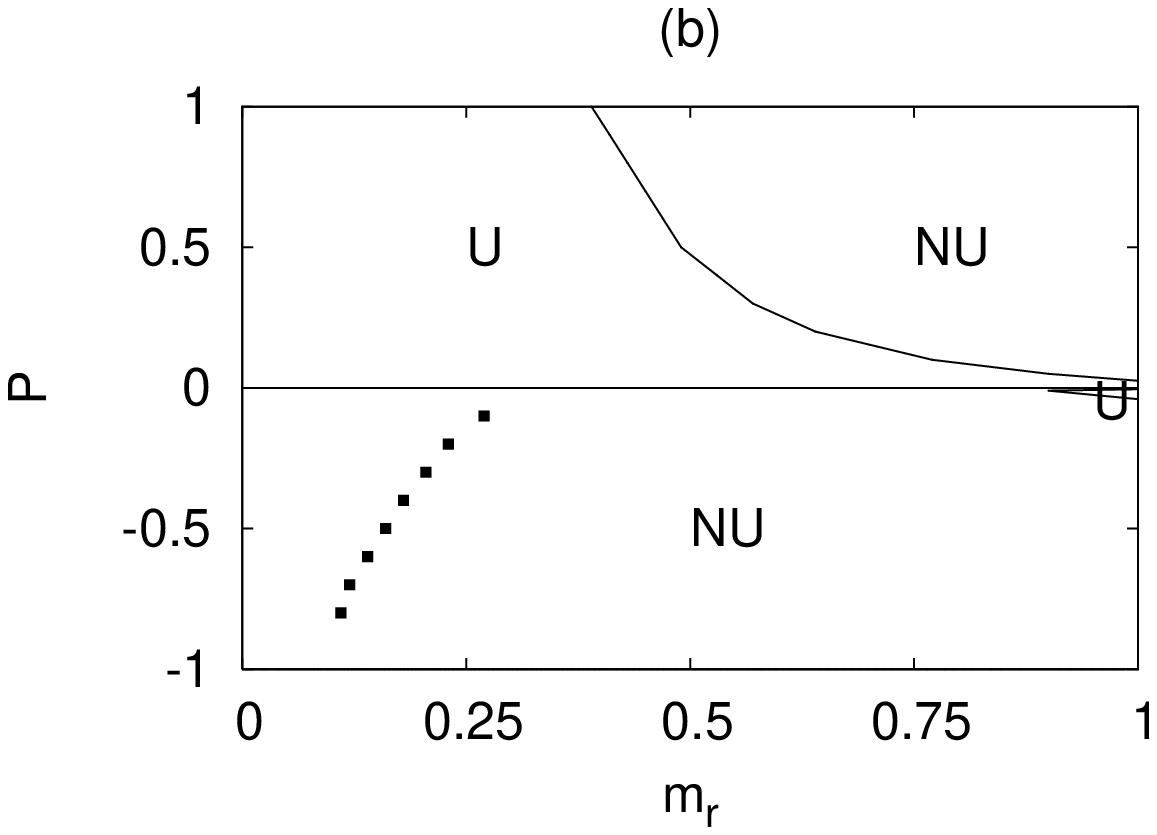} }}
\centerline{\scalebox{0.6}{\includegraphics{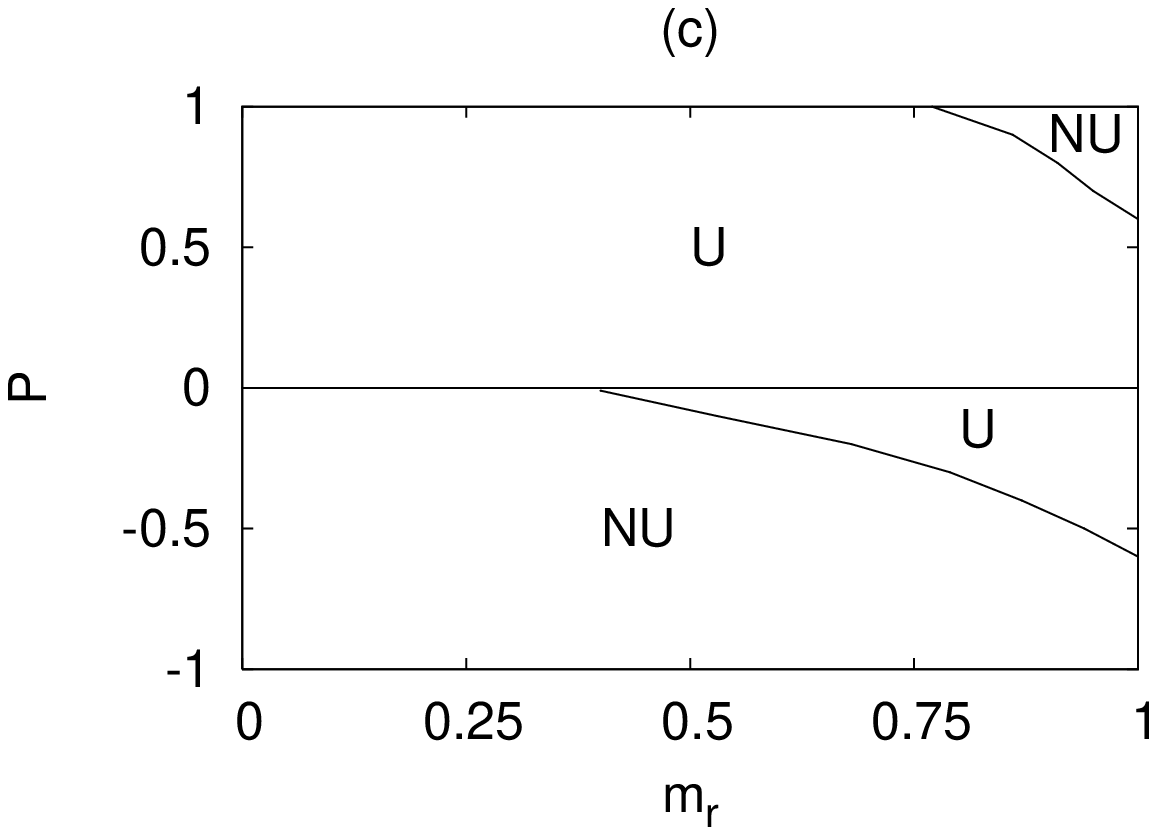} }}
\caption{\label{fig:phase.bec} 
Phase diagram of $P = (N_\uparrow - N_\downarrow)/(N_\uparrow + N_\downarrow)$ 
versus $m_r = m_\uparrow / m_\downarrow$ on the BEC side 
when a) $1/(k_{F,+} a_F) = 0.25$, b) $1/(k_{F,+} a_F) = 1$ and c) $1/(k_{F,+} a_F) = 2$.
We show normal (N), uniform (U) or non-uniform (NU) superfluid phases.
The dotted line (black squares) separate topologically distinct regions.
}
\end{figure}

Based on all the previous criteria, we construct the $P$ versus $m_r$ phase diagram
for seven sets of interaction strengths:
$1/(k_{F,+} a_F) = -2$, $-1$ and $-0.25$ on the BCS side shown in Fig.~\ref{fig:phase.bcs}; 
$1/(k_{F,+} a_F) = 0$ at unitarity shown in Fig.~\ref{fig:phase.unitarity}; and 
$1/(k_{F,+} a_F) = 0.25$, $1$ and $2$ on the BEC side shown in Fig.~\ref{fig:phase.bec}.
In these diagrams, the $\uparrow$ ($\downarrow$) label
always corresponds to lighter (heavier) mass such that lighter (heavier) fermions
are in excess when $P > 0$ $(P < 0)$. Notice that this choice spans all possible population 
imbalances and mass ratios. 

In Figs.~\ref{fig:phase.bcs}-\ref{fig:phase.bec},
we indicate the regions of normal (N), and uniform (U) or non-uniform (NU)
superfluid phases. The black squares indicate the transition line that separates topological phases I and II.
In all phase diagrams, phase I (II) always appears to the left (right) of the dotted lines for $P > 0$,
while phase I (II) always appears to the right (left) of the dotted lines for $P < 0$.

The normal phase is characterized by a vanishing order parameter ($\Delta_0 = 0$),
while the uniform superfluid phase is characterized by $\rho_0 (0) > 0$ and  
$\partial^2 \Omega_0 /\partial \Delta_0^2 > 0$.
The non-uniform superfluid phase is characterized by
$\rho_0 (0) < 0$ and/or $\partial^2 \Omega_0 /\partial \Delta_0^2 < 0$, and
it should be of the LOFF-type having one wavevector modulation only near the BCS limit~\cite{FF,LO}, 
although closer to unitarity, we expect the non-uniform phase to be substantially 
different from the LOFF phases either having spatial modulation that would encompass several 
wavevectors or presenting complete phase separation between paired and unpaired fermions. 
However, from numerical calculations, the superfluid density criterion seems to be weaker for all parameter 
space and the non-uniform superfluid phase is characterized by
$\partial^2 \Omega_0 /\partial \Delta_0^2 < 0$, which indicates possible phase separation,
since at least one of the eigenvalues of the compressibility matrix $\kappa_{\sigma, \sigma'}$ 
also becomes negative.
Therefore, for a homogeneous system paired (or superfluid) fermions and unpaired (or excess) fermions 
coexist in the uniform superfluid regions, they are phase separated in non-uniform superfluid regions,
and the topological transition from phase I to phase II may not be accessible. 
However, in a harmonic trap with a large superfluid region at the center, the topological phases should
be observable since the central region is essentially a ``uniform superfluid'' with the excess 
fermions at the edge~\cite{iskinmixture}. The peculiar momentum distribution of different topological phases 
would be smeared out by the trapping potential, but their marked signatures should still be present. 
Furthermore, these topological phases may be accessible in trapped systems at finite temperature~\cite{duan-qpt}, 
or in optical lattices~\cite{iskin-lattice}.

As shown in Fig.~\ref{fig:phase.bcs}(a) and~\ref{fig:phase.bcs}(b), we find a small region of uniform
superfluidity on the BCS side 
only when the mass anisotropy is small and the
lighter fermions are in excess ($P > 0$).
Thus, to probe the largest amount of phases on the BCS side, mixtures consisting of $^6$Li and $^{40}$K ($m_r \approx 0.15$) 
or $^6$Li and $^{87}$Sr ($m_r \approx 0.07$) are good candidates.
In the rest of the phase diagram, 
we find a quantum phase transition from the non-uniform 
superfluid to the normal phase beyond a critical population imbalance 
for both positive and negative $P$.
The phase space of uniform superfluidity expands while that of the 
normal phase shrinks with increasing interaction strength as shown 
in Figs.~\ref{fig:phase.bcs}(b) and~\ref{fig:phase.bcs}(c).

This general trend continues into the unitarity limit [$1/(k_{F,+} a_F) = 0$] 
as shown in Fig.~\ref{fig:phase.unitarity}. Since this limit is theoretically 
important as well as experimentally accessible, it is useful to analyze the phase
diagram as a function of population imbalance and mass anisotropy.
Notice that Fermi mixtures corresponding to mass ranges $0 < m_r < 0.23$,
like $^6$Li and $^{87}$Sr ($m_r \approx 0.07$) and $^6$Li and $^{40}$K ($m_r \approx 0.15$) 
have phase diagrams which are qualitatively different from those corresponding to mass
ratios $ 0.23 < m_r < 0.45$ like $^6$Li and $^{25}$Mg ($m_r \approx 0.24$) and $^6$Li and $^2$H ($m_r \approx 0.33$), 
since a topological transition line may be accessible in the second range.
Furthermore, only NU and N phases are accessible at unitarity 
for Fermi mixtures in the range of mass ratios $0.45 < m_r < 1$ like $^{40}$K and $^{87}$Sr ($m_r \approx 0.64$)
or any equal mass mixtures. 
Notice that our results for the case of equal masses ($m_r = 1$) are in close 
agreement with recent MIT experiments~\cite{mit} in a trap.
At unitarity, our non-uniform superfluid to normal state 
boundary occurs at $P \approx \pm 0.73$, and the MIT group obtains 
$P \approx \pm 0.70(4)$ for their superfluid to normal boundary.

Additional increase of interaction strength beyond unitarity on the BEC side 
leads to further expansion (shrinkage) of the uniform superfluid (normal) region
as shown in Fig.~\ref{fig:phase.bec}(a) and~\ref{fig:phase.bec}(b).
When heavier fermions are in excess ($P < 0$), a uniform superfluid phase 
is not possible for any mass anisotropy until a critical interaction
strength is reached. The critical interaction strength corresponds to
$1/(k_{F,+} a_F) \approx 0.8$ for $m_r = 1$.
Further increase of interaction strength towards the BEC limit [$1/(k_{F,+} a_F) > 1$],
leads to further expansion (shrinkage) of the uniform (non-uniform) superfluid
region as shown in Fig.~\ref{fig:phase.bec}(c), and
only the uniform superfluid phase exists in the extreme BEC limit
[$1/(k_{F,+} a_F) \gg 1$] even for $P < 0$ (not shown).

Having discussed the ground state phase diagrams, we present next fluctuation
effects beyond the saddle point approximation.

\section{Gaussian fluctuations}
\label{sec:fluctuations}

In this section, we discuss the (gaussian) fluctuation effects around 
the saddle point solutions at finite and zero temperatures.
Near the critical temperature ($T \approx T_c$)
we discuss the time-dependent Ginzburg-Landau (TDGL) equation, and
at zero temperature ($T = 0$) we analyse the collective phase (or Goldstone) 
mode as well as the effects of harmonic trap in the BEC limit, which
are discussed next.

\subsection{Time-dependent Ginzburg-Landau equation near critical temperatures}
\label{sec:finitetemperatures}

Our basic motivation here is to investigate the low frequency and long wavelength 
behavior of the order parameter near $T_c$
where $\Delta_0 = 0$, and derive the TDGL equation~\cite{carlos}.
We use the small $\mathbf{q}$ and $iv_{\ell} \to \omega + i\delta$ expansion of 
\begin{eqnarray*}
L^{-1}(q) = \frac{1}{g} - \sum_{\mathbf{k}} 
\frac{1 - n_f(\xi_{\mathbf{k} + \mathbf{q}/2,\uparrow}) - n_f(\xi_{\mathbf{k} - \mathbf{q}/2,\downarrow})} 
{\xi_{\mathbf{k} + \mathbf{q}/2,\uparrow} + \xi_{ \mathbf{k} - \mathbf{q}/2,\downarrow} - iv_\ell} |\Gamma_\mathbf{k}|^2,
\end{eqnarray*}
where $L^{-1}(q) = \mathbf{F}^{-1}_{1,1}(q)$, to obtain the
TDGL equation
\begin{equation}
\label{eqn:tdgl}
\left[ a + b|\Lambda(x)|^2 - \sum_{i,j}\frac{c_{ij}}{2}\nabla_i\nabla_j - 
id\frac{\partial}{\partial t} \right]\Lambda(x) = 0
\end{equation}
in the real space $x = (\mathbf{x},t)$ representation.
Here, $n_f(x) = 1/[\exp(\beta x) + 1]$ is the Fermi distribution.

Expressions for the coefficients $a$, $b$, $c_{ij}$ and $d$ are presented in appendix~\ref{sec:app.b}.
The condition $ a = 0 $ corresponds to the Thouless criterion, and the coefficient
of the nonlinear term is positive ($b >0$) guaranteeing
the stability of the effective theory. The kinetic energy coefficient
$c_{ij}$ is an effective inverse mass tensor which reduces to a scalar $c$ in the s-wave case.
The time-dependent coefficient $d$ is a complex number, and 
its imaginary part reflects the decay of Cooper pairs into the two-particle continuum 
for $\mu_+ > 0$. However, for $\mu_+ < 0$, the imaginary part of $d$
vanishes and the behavior of the order parameter $\Lambda(x)$ is
propagating reflecting the presence of stable (long lived) bound states.

Since a uniform superfluid phase is more stable in the BEC side, 
we calculate analytically all coefficients in the BEC limit
where $|\mu_\pm| \sim |\epsilon_b|/2 \gg T_{\rm c}$.
We obtain
$a = a_1 + a_2 = - V m_+^2 (2\mu_+ - \epsilon_b) a_F/(8\pi) + V m_+ n_e a_F^2$,
$b = b_1 + b_2 = V m_+^3 a_F^3/(16 \pi) - V m_+^2 (\partial n_e/\partial \mu_e) a_F^4$,
$c = V m_+^2 a_F / [8\pi(m_\uparrow + m_\downarrow)]$, and
$d = V m_+^2 a_F / (8\pi)$.
Here, $e$ labels the excess type of fermions and 
$n_e$ is the density of unpaired fermions.
Through the rescaling $\Psi(x) = \sqrt{d}\Lambda(x)$,
we obtain the equation of motion 
for a dilute mixture of weakly interacting bosons and fermions
\begin{eqnarray}
-\mu_B \Psi(x) &+& \left[U_{BB}|\Psi(x)|^2 + U_{BF} n_e(x) \right] \Psi(x) \nonumber \\ 
&-& \frac{\nabla^2 \Psi(x)}{2m_B}  - i\frac{\partial \Psi(x)}{\partial t} = 0,
\label{eqn:GL-BEC}
\end{eqnarray}
with bosonic chemical potential 
$
\mu_B = - a_1/d = 2\mu_+ - \epsilon_b,
$
mass 
$
m_B = d/c = m_\uparrow + m_\downarrow,
$
and repulsive boson-boson 
$
U_{BB} = V b_1/d^2 = 4\pi a_F / m_+
$
and boson-fermion
$
U_{BF} = V a_1/(d n_e) = 8\pi a_F / m_+
$
interactions.
This procedure also yields the spatial density of unpaired fermions given by
\begin{eqnarray}
n_e(x) &=& [a_2/d + b_2|\Psi(x)|^2/d^2]/U_{BF} \nonumber \\
&=& n_e - U_{BF} (\partial n_e/\partial \mu_e) |\Psi(x)|^2.
\end{eqnarray}
Since $\partial n_e/\partial \mu_e > 0$ the unpaired fermions avoid regions
where the boson field $|\Psi (x)|$ is large. Thus, in a harmonic trap, the bosons condense
at the center and the unpaired fermions tend to be at the edges.
Notice that, Eq.~(\ref{eqn:GL-BEC}) reduces to the Gross-Pitaevskii equation 
for equal masses with $P = 0$~\cite{carlos},
and to the equation of motion for equal masses with $P \ne 0$~\cite{pieri}.

Next, we recall the standard definitions of the interactions in terms of the scattering
lenghts $U_{BB} = 4\pi a_{BB}/m_B$ and $U_{BF} = 4\pi a_{BF}/m_{BF}$, where 
$m_B$ is the mass of the bosons and $m_{BF} = 2m_{e} m_B/(m_B + m_e)$ is twice
the reduced mass of a boson of mass $m_B$ and an excess fermion of mass $m_e$. Combining these definitions
with  our results for $U_{BB}$ and $U_{BF}$ in terms of the fermion-fermion scattering length $a_F$, we can 
directly relate the boson-boson scattering parameter
\begin{equation}
\label{eqn:abb}
a_{BB} = \frac{m_B}{m_+} a_F = \left[ 
1 + \frac{m_\uparrow}{2m_\downarrow} + \frac{m_\downarrow}{2m_\uparrow} 
\right] a_F
\end{equation}
and the boson-fermion scattering parameter 
\begin{equation}
\label{eqn:abf}
a_{BF} = \frac{4 m_B m_e}{m_+ (m_B + m_e)} a_F
\end{equation}
to $a_F$. Notice that these expressions reduce to 
$a_{BB} = 2a_F$ and $a_{BF} = 8a_F/3$ for equal masses~\cite{carlos,pieri}.
A better estimate for $a_{BB}$ can be found in the literature~\cite{gora}.

Since the effective boson-fermion system is weakly interacting, 
the BEC temperature is
$
T_c = 2\pi [n_B/\zeta(3/2)]^{2/3}/m_B,
$
where $\zeta(x)$ is the Zeta function and $n_B = (n - n_e)/2$.

Notice that, the effective total number equation for the boson-fermion mixture can be written as
\begin{equation}
N \approx \sum_{\mathbf{k}} n_f(\xi_{\mathbf{k},e}) 
+ 2\sum_{\mathbf{q}} n_b \left[ \frac{|\mathbf{q}|^2}{2m_B} - \widetilde{\mu}_B \right],
\end{equation}
where $n_b(\varepsilon) = 1/[\exp(\beta \varepsilon) - 1]$ is the Bose distribution
and $\widetilde{\mu}_B \to 0^-$ includes also the Hartree shift.
In the limit when $T_c \to 0$, we obtain the critical chemical potential 
for unpaired fermions at the normal-to-stable uniform superfluid boundary as given by 
$
\mu_e = 2^{2/3} (m_+/m_e) \epsilon_{F,+},
$
where $e = (\uparrow,\downarrow)$ labels excess type of atoms.
Since, $\mu_+ \to \epsilon_b/2$ in this limit, the critical chemical potential imbalance
is given by
$
\mu_- = \gamma_e [- \epsilon_b/2 + 2^{2/3} (m_+/m_e) \epsilon_{F,+}],
$
where $\epsilon_b = - 1/(m_+ a_F^2)$ is the binding energy, and
$\gamma_\uparrow = + 1$ and $\gamma_\downarrow = -1$.

This concludes our analysis for the homogenous mixture of two types of ultra-cold fermions
at finite temperatures. Next we discuss collective excitations at zero temperature.

\subsection{Sound velocity at zero temperature}
\label{sec:sound-velocity}

In order to obtain the collective mode spectrum, we 
use the effective action defined in Eq.~(\ref{eqn:eff-action})
and express $\Lambda(q) = [ \lambda(q) +i \theta(q) ] / \sqrt{2}$
in terms of the amplitude $\lambda(q)$ and phase $\theta(q)$ fields,
respectively.
Using the matrix elements of $\mathbf{F}^{-1}$ defined in Eqs.~(\ref{eqn:f1-matrix})
and~(\ref{eqn:f2-matrix}) and described in appendix~\ref{sec:app.a}, we can obtain the 
matrix elements of the fluctuation matrix in the rotated basis $(\lambda(q),\theta(q))$.
The diagonal elements of the fluctuation matrix in the rotated basis become
$
\mathbf{M}^{-1}_{\lambda,\lambda}	(q) = 
[\mathbf{F}^{-1}_{1,1} + \mathbf{F}^{-1}_{1,2} + \mathbf{F}^{-1}_{2,1} + \mathbf{F}^{-1}_{2,2}]/2
$
and
$
\mathbf{M}^{-1}_{\theta,\theta}(q) = 
[\mathbf{F}^{-1}_{1,1} - \mathbf{F}^{-1}_{1,2} - \mathbf{F}^{-1}_{2,1} + \mathbf{F}^{-1}_{2,2}]/2,
$
while the off-diagonal elements become
$
\mathbf{M}^{-1}_{\lambda,\theta}(q) = 
i [\mathbf{F}^{-1}_{1,1} - \mathbf{F}^{-1}_{1,2} + \mathbf{F}^{-1}_{2,1} - \mathbf{F}^{-1}_{2,2}]/2
$
with
$
\mathbf{M}^{-1}_{\theta,\lambda}(q) = (\mathbf{M}^{-1}_{\lambda,\theta})^*(q).
$

The collective modes are found from the poles of the fluctuation  matrix $\mathbf{M} (q)$ determined
by the condition $\det \mathbf{M}^{-1} (q) = 0$, when the usual analytic continuation
$iv_\ell \rightarrow w + i0^+$ is performed.
The easiest way to get the phase collective modes is to integrate out the 
amplitude fields to obtain a phase-only effective action.
To obtain the long wavelength dispersions for the collective modes at $T = 0$, 
we consider $|P| \to 0$ or $k_{F,+} = k_{F,\uparrow} = k_{F,\downarrow}$ limit,
and expand the matrix elements of $\mathbf{F}^{-1}(q)$ 
to second order in $|\mathbf{q}|$ and $w$ to get
$
\mathbf{M}_{\lambda,\lambda}^{-1} (q) = A + C |\mathbf{q}|^2 - D w^2,
$
$
\mathbf{M}_{\theta,\theta}^{-1} (q) = Q |\mathbf{q}|^2 - R w^2
$
and
$
\mathbf{M}_{\lambda,\theta}^{-1} (q) = i B w,
$
such that
\begin{eqnarray*}
\mathbf{M}^{-1}(\mathbf{q},w) = \left( \begin{array}{cc} A + C|\mathbf{q}|^2 - Dw^2 & iBw 
\\ -iBw & Q|\mathbf{q}|^2 - Rw^2 \end{array}\right).
\label{eqn:coll}
\end{eqnarray*}
The expansion coefficients are given in the Appendix~\ref{sec:app.b}.
Thus, there are two branches for the collective excitations, but we focus only on
the lowest energy one correponding to the Goldstone mode with dispersion
$
w(\mathbf{q}) = v |\mathbf{q}|,
$
where 
\begin{equation}
v = \sqrt{ \frac{A Q}{AR + B^2} }
\end{equation}
is the speed of sound. Extra care is required when $P \ne 0$ since Landau damping causes collective excitations 
to decay into the two-quasiparticle continuum even for the s-wave case, since gapless fermionic (quasiparticle)
excitations are present (see Fig.~\ref{fig:qpt}).

The BCS limit is characterized by the criteria $\mu_+ > 0$ and 
$\mu_+ \approx \epsilon_{F,+} \gg |\Delta_0|$.
The expansion of the matrix elements to order $|{\mathbf q}|^2$ and $w^2$ is
performed under the condition $[w,|\mathbf{q}|^2/(2m_+)] \ll |\Delta_0|$.
The coefficient that couples phase and amplitude fields
vanish ($B = 0$) in this limit. Thus, there
is no mixing between the phase and amplitude modes.
The zeroth order coefficient is
$
A = {\cal D} ,
$
and the second order coefficients are
$
C = Q/3 = {\cal D} v_{F,\uparrow} v_{F,\downarrow}/ (36|\Delta_0|^2),
$
and
$
D = R/3 = {\cal D} / (12|\Delta_0|^2).
$
Here, $v_{F,\sigma} = k_{F,\sigma}/m_\sigma $ is the Fermi velocity and 
${\cal D} = m_+ V k_{F,+}/(2\pi^2)$ is the 
density of states per spin at the Fermi energy.
Thus, we obtain
\begin{equation}
v = \sqrt{ \frac{v_{F,\uparrow} v_{F,\downarrow}}{3} } = \sqrt{v_\uparrow v_\downarrow},
\end{equation}
with $v_\sigma = v_{F,\sigma}/\sqrt{3}$, 
which reduces to the Anderson-Bogoliubov relation when the masses are equal.

\begin{figure} [htb]
\centerline{\scalebox{0.6}{\includegraphics{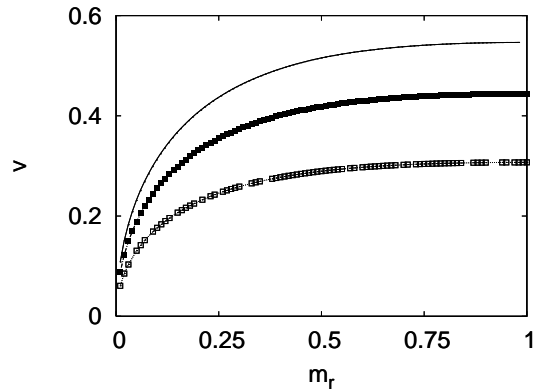} }}
\caption{\label{fig:collective}
Sound velocity $v$ (in units of $v_{F,+} = k_{F,+}/m_+$) versus $m_r$ for
$1/(k_{F,+} a_F) = -1$ (solid line),
$1/(k_{F,+} a_F) = 0$ (solid squares) and 
$1/(k_{F,+} a_F) = 1$ (hollow squares).
Here, populations are equal ($P = 0$).
}
\end{figure}

On the other hand, the BEC limit is characterized by the criteria $\mu_+ < 0$ and 
$\xi_{\mathbf{k},+} \gg |\Delta_0|$.
The expansion of the matrix elements to order $|{\mathbf q}|^2$ and $w^2$ is
performed under the condition $[w, |\mathbf{q}|^2/(2m_+)] \ll |\mu_+|$.
The coefficient $B \ne 0$ indicates that the amplitude and phase fields are mixed.
The zeroth order coefficient is
$
A = \kappa |\Delta_0|^2 / (2|\mu_+|),
$
the first order coefficient is
$
B = \kappa,
$
and the second order coefficients are
$
C = Q = \kappa/[2(m_\uparrow + m_\downarrow)]
$
and
$
D = R = \kappa /(8|\mu_+|),
$
where
$
\kappa = {\cal D} / (32\sqrt{|\mu_+|\epsilon_{F,+}}).
$
Thus, we obtain
\begin{equation}
v = \frac{|\Delta_0|}{\sqrt{4(m_\uparrow + m_\downarrow) |\mu_+|}}
= \sqrt{v_\uparrow v_\downarrow}
\end{equation}
with $v_\sigma = \sqrt{2\pi n_\sigma a_F / m_\sigma^2}$.
Notice that the sound velocity is very small and its smallness is controlled
by the scattering length $a_F$.
Furthermore, in the theory of weakly interacting dilute Bose gas, 
the sound velocity is given by 
$
v_B = \sqrt{4\pi a_{BB} n_B / m_B^2}.
$
Making the identification that the density of 
pairs is $n_B = n_+$, the mass of the bound pairs is 
$m_B = m_\uparrow + m_\downarrow$ and that the Bose scattering length is 
\begin{equation}
a_{BB} = \frac{m_B}{m_+} a_F = \left[
1 + \frac{m_\uparrow}{2m_\downarrow} + \frac{m_\downarrow}{2m_\uparrow}
\right] a_F, 
\end{equation}
$v_B$ reduces to the well known Bogoliubov relation when the masses are equal.
Therefore, the strongly interacting Fermi gas with two species can be described as 
a weakly interacting Bose gas at zero temperature as well as at finite temperatures~\cite{iskin-mixture}.
Notice that $a_{BB}$ reduces to $a_{BB} = 2a_F$ for equal masses~\cite{jan} in the
Born approximation. A better estimate for $a_{BB}$ can be found in the literature~\cite{gora}.

In Fig.~\ref{fig:collective}, we show the sound velocity as a function of the
mass ratio $m_r$ for three values of the scattering parameter $1/(k_{F,+} a_F) = -1, 0$ and $1$ corresponding
to the BCS side [$1/(k_{F,+} a_F) = -1$], unitarity [$1/(k_{F,+} a_F) = 0$], and to the BEC side 
[$1/(k_{F,+} a_F) = 1$]. Notice that the speed of sound could be measured for a given $m_r$ 
using similar techniques as in the single species case $m_r = 1$~\cite{thomas,grimm}.

This concludes our analysis for collective excitations in 
the homogenous mixture of a two types of fermions at zero temperature. 
We discuss next the effective field theory between paired and unpaired
fermions.

\subsection{Weakly interacting paired and excess fermions (Bose-Fermi mixtures) at zero temperature}
\label{sec:paired-unpaired-fermions}

In this section, we concentrate on the BEC regime, where the paired and
unpaired (excess) fermions can be described by a mixture of molecular bosons
and fermions. In this limit, the resulting equation of motion is identical
to Eq.~(\ref{eqn:GL-BEC}) near the critical temperature, except that all parameters
are evaluated at zero temperature. Thus, at low temperatures the system
continues to behave as a dilute mixture of weakly interacting bosons 
(formed from paired fermions) and unpaired fermions, and can be described by
the free energy density
\begin{equation}
{\cal F}(\mathbf{x}) = E (\mathbf{x}) - \mu_e n_e (\mathbf{x}) - \mu_{B} |\Psi (\mathbf{x})|^2,
\end{equation}
where the energy density is 
%
$$
E (\mathbf{x}) =  K_B + K_F + \frac{1}{2} U_{BB} |\Psi (\mathbf{x})|^4  + 
U_{BF} n_e (\mathbf{x}) |\Psi (\mathbf{x})|^2.
$$
%
Here, $K_B$ is the kinetic energy density of bosons (assumed to be much smaller than all the
other energies) and $K_F$ is the kinetic energy density of fermions. Averaging these energy densities 
over the spatial coordinates $F = \int d\mathbf{x} {\cal F}(\mathbf{x})/V$ leads to a ground state average free
energy density
%
$$
F = \frac{1}{2} U_{BB} n_B^2  + U_{BF} n_e n_B + \frac{3}{5} \epsilon_{F,e} n_e - \mu_e n_e - \mu_{B} n_B,
$$
%
where $n_e$ ($n_B$) is the average density of fermions (bosons), and
$\epsilon_{F,e}$ is the fermi energy of the excess fermions.
Using the positivity of the Bose-Fermi compressibility matrix 
$\kappa_{\alpha, \beta} = \partial \mu_{\alpha}/\partial n_{\beta}$, where $\alpha, \beta = e, B$, 
one can show that bosons and fermions phase separate when the condition
\begin{equation}
n_e \ge \frac{4\pi}{3} \left(\frac{\pi}{m_e}\right)^3 \left(\frac{U_{BB}}{U_{BF}^2}\right)^3
\end{equation}
is satisfied~\cite{viverit}.

Using the boson-boson and boson-fermion interactions
$U_{BB} = 4\pi a_{BB}/m_B$ and $U_{BF} = 4\pi a_{BF}/m_{BF}$, the scattering
paramaters indicated in Eqs.~(\ref{eqn:abb}) and~(\ref{eqn:abf}), as well as the relations
$|P| = n_e/n$, $n = n_{\uparrow} + n_{\downarrow} = k_{F,+}^3/(3\pi^2)$
and $n_B = (n-n_e)/2$, then phase separation occurs when
\begin{equation}
|P| \ge \frac{1}{2} \left(\frac{\pi}{8}\right)^3 \left(\frac{m_+/m_e}{k_{F,+} a_F}\right)^3.
\label{eqn:psc}
\end{equation}
Here $e$ labels excess type of fermions, and $m_e$ is the mass of the unpaired fermions,
and $m_+$ is twice the reduced mass of the $\uparrow$ and $\downarrow$ fermions. 
This expression is quantitatively correct in its region of validity, i.e., when
$1/(k_{F,+} a_F) \gg 1$, however, it still gives semi-quantitative results for 
$1/(k_{F,+} a_F) \gtrsim 2$. For instance, in the case of an equal mass mixture, this expression would suggest 
that the resulting Bose-Fermi mixture is uniform when $1/(k_{F,+}a_F) > 2.5$ for $|P| \to 0.5$, and 
when $1/(k_{F,+}a_F) > 3.2$ for $|P| \to 1$.  
From our numerical calculations we find $1/(k_{F,+}a_F) > 1.9$ for $|P| \to 0.5$, and 
when $1/(k_{F,+}a_F) > 2.4$ for $|P| \to 1$.

The analytic expression given in Eq.~(\ref{eqn:psc}) may be used as a guide for the boundary between
phase separation (non-uniform) and the mixed phase (uniform) for any mixture of fermions. 
In particular, this relation serves as an estimator for the phase bounday for future experiments
performed in the BEC limit of unequal mass fermions with population imbalance. This relation can 
be rewritten in terms of the mass ratio $m_r = m_{\uparrow}/m_{\downarrow}$ by realizing that the
ratio $m_+/m_e = 2m_r/(1 + m_r)$ when $\downarrow$ (heavier) fermions are in excess, and
that $m_+/m_e = 2/(1 + m_r)$ when $\uparrow$ (lighter) fermions are in excess. 
Thus, when $\downarrow$ (heavier) fermions are in excess, the critical polarization below which 
phase separation (non-uniform phase) occurs is 
\begin{equation}
P_{c,\downarrow}^{(1)} = - \frac{1}{2} 
\left(\frac{\pi}{4k_{F,+} a_F}\right)^3 
\left(\frac{m_r}{1 + m_r}\right)^3,
\label{eqn:psdown1}
\end{equation}
while, when $\uparrow$ (lighter) fermions are in excess, the critical polarization above which 
phase separation (non-uniform phase) occurs is 
\begin{equation}
P_{c,\uparrow}^{(1)} = + \frac{1}{2} 
\left(\frac{\pi}{4k_{F,+} a_F}\right)^3 
\left(\frac{1}{1 + m_r}\right)^3.
\label{eqn:psup1}
\end{equation}
Notice that in the equal mass ($m_r = 1$) case $P_{c,\downarrow}^{(1)} = - P_{c,\uparrow}^{(1)}$
as required by symmetry.

\begin{figure} [htb]
\centerline{\scalebox{0.6}{\includegraphics{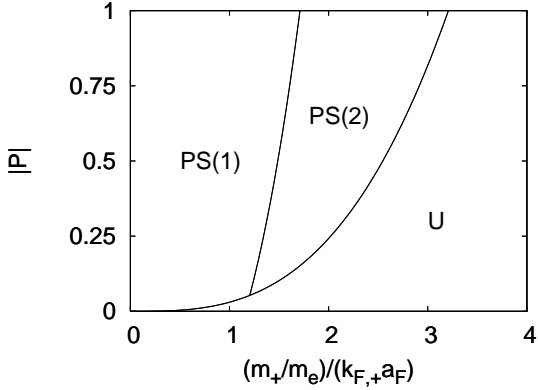} }}
\caption{\label{fig:viverit} 
Phase diagram of $|P| = |N_\uparrow - N_\downarrow|/(N_\uparrow + N_\downarrow)$ 
versus $(m_+/m_e)/(k_{F,+}a_F)$ in the BEC limit.
We show the uniform superfluid (U) phase where paired and unpaired fermions
coexist, and the phase separated non-uniform superfluid phases PS(1) and PS(2).
The PS(1) region labels phase separation between pure unpaired (excess) and pure tightly paired 
fermions (bosons), while the PS(2) region labels phase separation between 
pure unpaired (excess) fermions and a mixture of unpaired and tightly paired fermions.
}
\end{figure}

In addition, we can also describe analytically a finer structure of non-uniform 
(phase separated) superfluid phases deep into the BEC regime. 
For a weakly interacting Bose-Fermi mixture, the phase separated region consists two regions:
PS(1), where there is phase separation between pure fermions and pure bosons (tightly paired fermions),
and PS(2), where there is phase separation between pure fermions and a mixture of fermions and bosons (tighly
bound fermions). Following the method of Ref.~\cite{viverit}, we obtain analytically the condition
\begin{equation}
n_e \ge \frac{1125 \pi^4}{128 m_e^4} \frac{U_{BB}^3}{U_{BF}^6} - \frac{5}{8}n_B,
\end{equation}
for the transition from the PS(2) to the PS(1) phase. 

Using our effective boson-boson ($U_{BB}$) and effective boson-fermion $U_{BF}$ interactions, we can 
rewrite this relation as 
\begin{equation}
|P| \ge \frac{8}{11}\left(\frac{15\pi}{64}\right)^3\left(\frac{m_+/m_e}{k_{F,+} a_F}\right)^3 - \frac{5}{11},
\label{eqn:psc2}
\end{equation}
where we used $n_B = (n - n_e)/2$ as the boson density and $|P| = n_e/n$.
These phase boundaries can also be expressed in terms of $m_r$ and $1/(k_{F,+}a_F)$ as follows.
When $\downarrow$ (heavier) fermions are in excess, the critical polarization below which 
the transition from PS(2) to PS(1) occurs is 
\begin{equation}
P_{c,\downarrow}^{(2)} = - \frac{1}{11} 
\left(\frac{15\pi}{16k_{F,+} a_F}\right)^3 
\left(\frac{m_r}{1 + m_r}\right)^3 + \frac{5}{11},
\label{eqn:psdown2}
\end{equation}
while, when $\uparrow$ (lighter) fermions are in excess, the critical polarization above which 
the transition from PS(2) to PS(1) occurs is 
\begin{equation}
P_{c,\uparrow}^{(2)} = + \frac{1}{11} 
\left(\frac{15\pi}{16k_{F,+} a_F}\right)^3 
\left(\frac{1}{1 + m_r}\right)^3 - \frac{5}{11}.
\label{eqn:psup2}
\end{equation}
Notice that in the equal mass ($m_r = 1$) case $P_{c,\downarrow}^{(2)} = - P_{c,\uparrow}^{(2)}$
as required by symmetry.

In Fig.~\ref{fig:viverit}, we show phase diagram of uniform and 
non-uniform superfluidity as a function of population imbalance $|P|$ 
and $(m_+/m_e)/(k_{F,+}a_F)$, which is strictly valid in the BEC 
limit when $1/(k_{F,+} a_F) \gg  1$.
In this figure, we show the uniform superfluid (U) phase where tightly paired and unpaired fermions
coexist, and phase separated (non-uniform) superfluid (PS) phases.
The PS(1) region labels phase separation between pure unpaired (excess) and pure tightly paired 
fermions (bosons), while the PS(2) region labels phase separation between 
pure unpaired (excess) fermions and a mixture of unpaired and tightly paired fermions.
The phase boundary between U and PS(2) phases is determined from Eq.~(\ref{eqn:psc}), 
and the phase boundary between PS(2) and PS(1) phases is determined from Eq.~(\ref{eqn:psc2}).
For a fixed mass anisotropy $m_r$, when $|P|$ is large, 
we find phase transitions from PS(1) to PS(2) to U phase as 
the interaction strength $1/(k_{F,+} a_F)$ increases.
However, when $|P|$ is small, we find a phase transition from the PS(1) to the U phase as 
$1/(k_{F,+} a_F)$ increases. This last result is unphysical and signals
the breakdown of the approximations.

\begin{figure} [htb]
\centerline{\scalebox{0.6}{\includegraphics{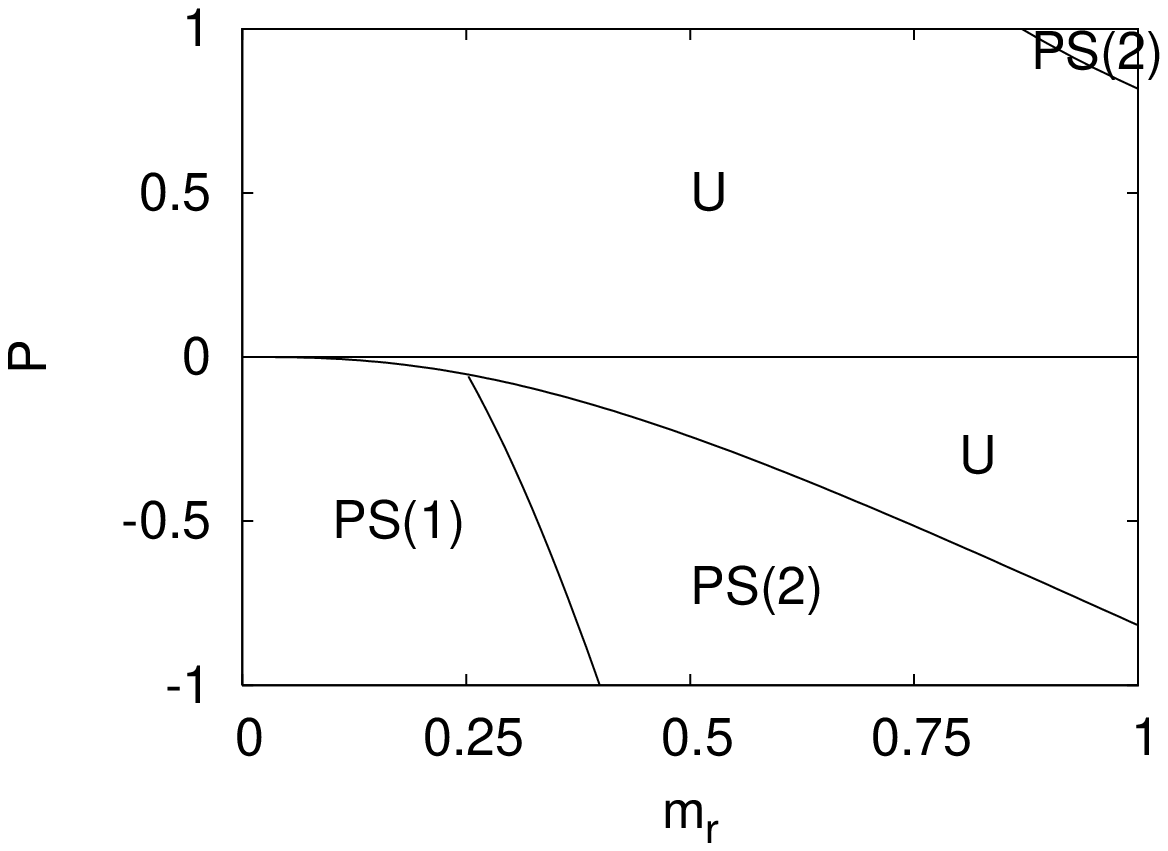} }}
\centerline{\scalebox{0.6}{\includegraphics{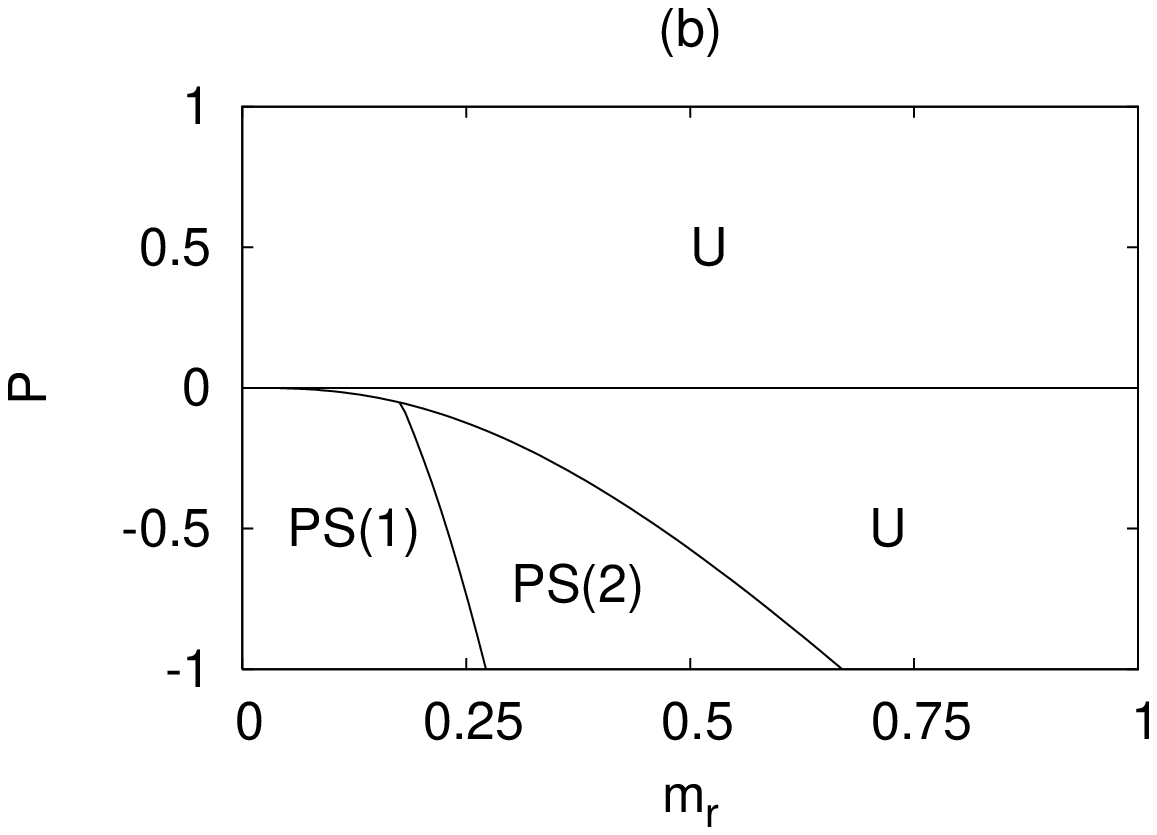} }}
\caption{\label{fig:BFphase} 
Phase diagram of $P = (N_\uparrow - N_\downarrow)/(N_\uparrow + N_\downarrow)$ 
versus $m_r = m_\uparrow / m_\downarrow$ in the BEC limit 
when (a) $1/(k_{F,+} a_F) = 3$ and (b) $1/(k_{F,+} a_F) = 4$.
We show the uniform superfluid (U) phase where paired and unpaired fermion
coexist, and phase separated non-uniform superfluid phases PS(1) and PS(2).
}
\end{figure}

In Fig.~\ref{fig:BFphase}, we show the phase diagram of population imbalance
$P$ versus mass anisotropy $m_r = m_\uparrow / m_\downarrow$ in the BEC limit 
when (a) $1/(k_{F,+} a_F) = 3$ and (b) $1/(k_{F,+} a_F) = 4$.
We indicate the uniform superfluid (U) phase where paired and unpaired fermion
coexist, and phase separated non-uniform superfluid phases PS(1) and PS(2).
The phase boundary between U and PS(2) phases is determined from 
Eq.~(\ref{eqn:psdown1}) when $P < 0$, and from Eq.~(\ref{eqn:psup1}) when $P > 0$.
In addition, the phase boundary between PS(2) and PS(1) phases is determined from
Eq.~(\ref{eqn:psdown2}) when $P < 0$, and PS(1) phase does not exist when $P > 0$.
Notice that these phase diagrams are very similar to 
the one given in Fig.~\ref{fig:phase.bec}(c),
with the added refinement of the non-uniform superfluid phases PS(1) and PS(2).
For a fixed interaction strength $1/(k_{F,+}a_F)$, when $|P|$ is large, 
we find phase transitions from PS(1) to PS(2) to U phase as 
the mass anisotropy $m_r$ increases.
However, when $|P|$ is small, we find a phase transition from the PS(1) to the U phase
as $m_r$ increases. This last result is unphysical and again 
signals the breakdown of the approximations.

To summarize, we analyzed analytically the structure of non-uniform 
(phase separated) superfluid phases in the BEC regime. 
However to understand the ultracold atomic experiments, one needs also to 
consider the trapping potential, which is discussed next.

\subsection{Effects of trapping potential}
\label{sec:trap}

For simplicity, we approximate the trapping potential by an isotropic harmonic function
where the potential energy is $V_\sigma(\mathbf{x}) = \alpha_\sigma |\mathbf{x}|^2/2$ such that
the local chemical potentials are given by 
\begin{equation}
\mu_\sigma(\mathbf{x}) = \mu_\sigma - \frac{1}{2} \alpha_\sigma |\mathbf{x}|^2.
\end{equation}
Here, $\alpha_\sigma$ is proportional to the trapping frequency of the $\sigma$ type fermion,
which is typically different for each kind of atom. In general, it is quite difficult to make 
completely isotropic traps and harmonic traps are typically elongated such that 
$V_\sigma(\mathbf{x}) = \alpha_{\sigma,x} x^2/2 + \alpha_{\sigma,y} y^2/2 + \alpha_{\sigma,z} z^2/2$ 
with $\alpha_{\sigma,x} = \alpha_{\sigma,y} \gg \alpha_{\sigma,z}$. However, the same qualitative behavior
occurs in the elongated or spherically symmetric (isotropic) traps, and we will confine ourselves for simplicity
to the isotropic case. When experimental data becomes available and all the numbers are known, one 
can revisit this problem for detailed comparison between theory and experiment.

\begin{figure} [htb]
\centerline{\scalebox{0.6}{\includegraphics{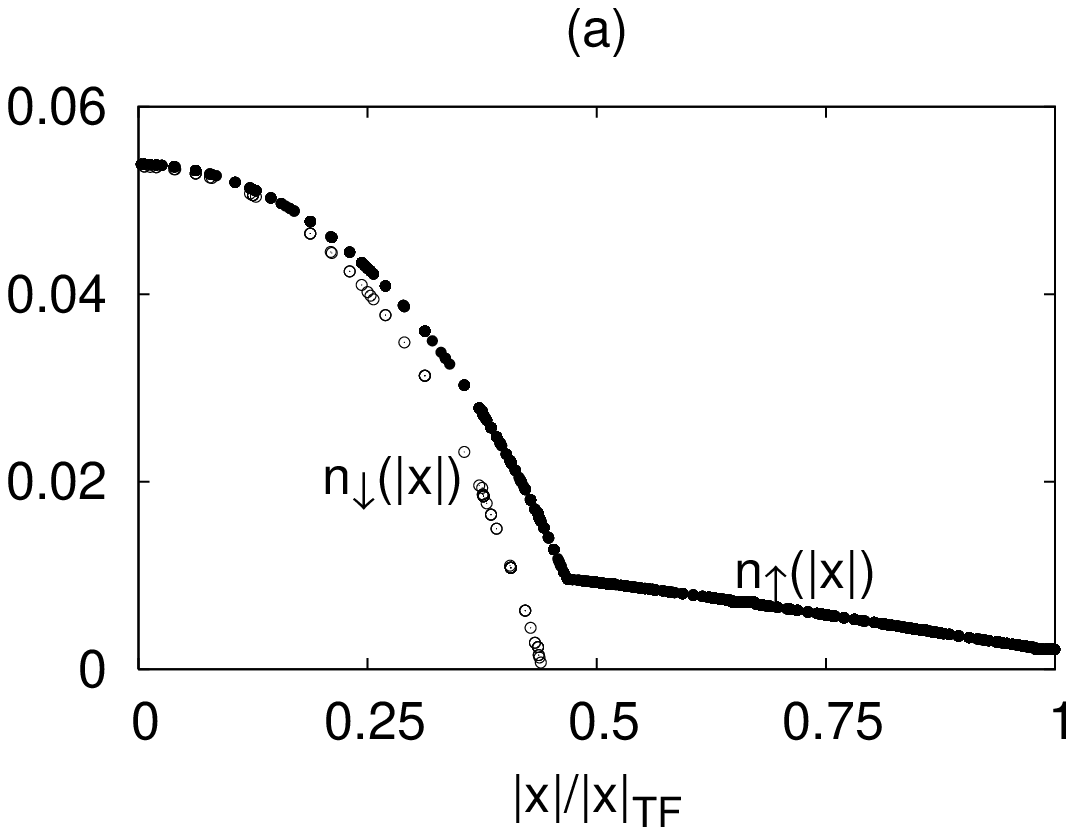}}}
\centerline{\scalebox{0.6}{\includegraphics{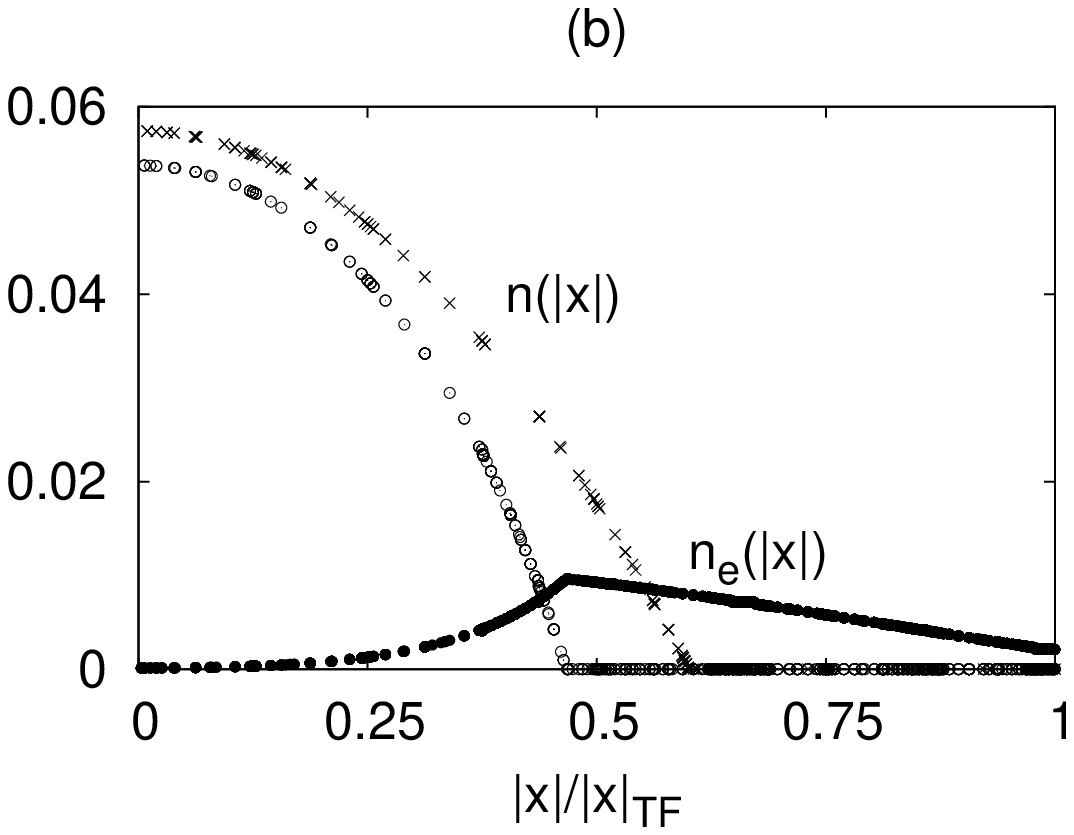}}}
\caption{\label{fig:trap1}
(a) Density $n_\sigma(|\mathbf{x}|)$ of fermions (in units of $k_{F,+}^3$), and
(b) total density of fermions $n(|\mathbf{x}|)$ (hollow circles) and density of 
unpaired fermions $n_e(|\mathbf{x}|)$ (solid circles) 
versus trap radius $|\mathbf{x}|/|\mathbf{x}|_{TF}$.
Here $P = 0.5$ and $1/(k_{F,+} a_F) = 2$. 
In (b), we also compare $n(|\mathbf{x}|)$ for $P = 0$ when $1/(k_{F,+} a_F) = 2$ (crosses).
}
\end{figure}

Again, we confine our discussion to the BEC regime, and obtain the equation of motion for a dilute
mixture of weakly interacting bosons and fermions at zero temperature
\begin{eqnarray}
&&-\mu_B \Psi(x) + \left[U_{BB}|\Psi(x)|^2 + U_{BF} n_e(x) \right] \Psi(x) + \nonumber \\ 
&& [V_\uparrow(\mathbf{x}) + V_\downarrow(\mathbf{x})]\Psi(x) - \frac{\nabla^2 \Psi(x)}{2m_B}  
= i\frac{\partial \Psi(x)}{\partial t},
\label{eqn:GL-BEC-t}
\end{eqnarray}
where the spatial density of unpaired fermions is 
\begin{eqnarray}
n_e(x) = n_{ex}(\mathbf{x}) - U_{BF} \frac{\partial n_{ex}(\mathbf{x})}{\partial \mu_e} |\Psi(x)|^2.
\end{eqnarray}
These results are quite similar to the case of equal masses~\cite{pieri}. Notice that 
setting $V_{\sigma} = 0$ reduces the problem to the free space case discussed in the 
previous subsection.
Here, 
$
n_{e}(\mathbf{x}) = (1/V) \sum_{\mathbf{k}} n_f \left[ \epsilon_{\mathbf{k}, e} - \mu_e(\mathbf{x}) \right],
$
where $n_f(\varepsilon) = 1/[\exp(\beta \varepsilon) + 1]$ is the Fermi distribution.
In the BEC limit when $a_F \to 0^+$, we can approximate the local density of unpaired fermions as
\begin{eqnarray}
n_e(x) \approx \frac{1}{V} \sum_{\mathbf{k}} n_f\left[\epsilon_{\mathbf{k}, e}
- \mu_e(\mathbf{x}) + U_{BF}  |\Psi(x)|^2 \right],
\end{eqnarray}
which at zero temperature leads to
\begin{eqnarray}
n_e(\mathbf{x}) = \frac{1}{6\pi^2} \lbrace 2m_e\left[\mu_e(\mathbf{x}) - U_{BF} n_B(\mathbf{x})\right] \rbrace^{3/2}.
\end{eqnarray}
Notice that the density of bosons at zero temperature 
is given by $n_B(\mathbf{x}) = |\Psi(\mathbf{x})|^2$.
Therefore, we need to solve Eq.~(\ref{eqn:GL-BEC-t}) self-consistently 
with the number of unpaired (excess)
$
N_e = \int d\mathbf{x} n_e(\mathbf{x})
$
and paired (bound)
$
N_{bf} =  2\int d\mathbf{x} n_B(\mathbf{x})
$
fermions such that the total number of fermions is $N = N_e + N_{bf}$.

Next we solve the self-consistency equations for a $^6$Li and $^{40}$K mixture ($m_r = 0.15$)
within the Thomas-Fermi (TF) approximation, where the kinetic energy term 
in Eq.~(\ref{eqn:GL-BEC-t}) is neglected. This leads to a coupled equation for
density of paired and unpaired fermions
\begin{eqnarray}
n_B(\mathbf{x}) \approx \frac{\mu_B - V_\uparrow(\mathbf{x}) - V_\downarrow(\mathbf{x}) - U_{BF} n_e(\mathbf{x})} {U_{BB}}.
\end{eqnarray}
In our numerical analysis, we choose for convenience 
$\alpha_\uparrow = \alpha_\downarrow = \alpha$, $1/(k_{F,+} a_F) = 2$ 
and $P = 0.5$ such that $N_\uparrow = 3N_\downarrow$. However, in a more realistic case 
$\alpha_\uparrow \ne \alpha_\downarrow = \alpha$, since atoms with different masses may experience
different trapping potentials due to their different polarizabilities.
  
In Fig.~\ref{fig:trap1}(a), we show the density $n_\sigma(|\mathbf{x}|)$ of $\sigma$ type fermion 
(in units of $k_{F,+}^3$) as a function of $|\mathbf{x}|/|\mathbf{x}|_{TF}$, where
$|\mathbf{x}|_{TF}$ is the TF radius defined by 
$\epsilon_{F,+} = k_{F,+}^2/(2m_+) = \alpha |\mathbf{x}|_{TF}^2/2$.
We also scale the total number of fermions with  $N = k_{F,+}^3 |\mathbf{x}|_{TF}^3/24$.
We find that the density of $\uparrow$ and $\downarrow$ fermions are 
similar close to the center of the trapping potential, while most of the excess
fermions are close to the edges.
In Fig.~\ref{fig:trap1}(b), we show the total density 
$n(|\mathbf{x}|) = n_\uparrow(|\mathbf{x}|) + n_\downarrow(|\mathbf{x}|)$ 
as well as unpaired fermions $n_e(|\mathbf{x}|) = n_\uparrow(|\mathbf{x}|) - n_\downarrow(|\mathbf{x}|)$.
In both figures, we find a clear indication of phase separation between paired and unpaired fermions.
In Fig.~\ref{fig:trap1}(b), we also compare the total density of fermions $n(|\mathbf{x}|)$ 
for the same parameters when the populations are balanced $N_\uparrow = N_\downarrow$.
When $P \ne 0$, the total local density of fermions at the center of the trap is
reduced in comparison to the $P = 0$ case for the same fermion scattering parameter, 
since the unpaired fermions are pushed away from the center of the trap due to $U_{BF}$.
These findings for unequal masses are similar to previous results on
equal mass mixtures~\cite{pieri,torma,yi,chevy,silva,haque}.

Lastly, for the parameters used, the harmonic trap tends to favor phase separation into a PS(1)-type 
phase where one has almost pure fermion and almost pure boson regions. In a harmonic trap
it may be still possible to realize the PS(2)-type phase where one has an almost pure fermion region
and an almost pure mixed phase of bosons and fermions, provided that good control over the trapping
potentials is possible. Having concluded our discussion of the effects of a trapping potential, we present
next the summary of our conclusions.

\section{Conclusions}
\label{sec:conclusions}

In summary, we analyzed the phase diagram of ultra cold mixtures of
two types of fermions (e.g., $^6$Li and $^{40}$K; $^6$Li and $^{87}$Sr; or $^{40}$K and $^{87}$Sr) 
from the BCS to the BEC limit as a function of scattering parameter, population imbalance, and mass anisotropy.
We found that the zero temperature phase diagram of population imbalance versus scattering parameter 
is asymmetric for unequal masses, having a larger stability region for uniform superfluidity 
when the lighter fermions are in excess. This result is in sharp 
contrast with the symmetric phase diagram for equal masses.

In addition, we discussed topological quantum phase transitions
associated with the disappearance or appearance of momentum space regions of zero quasiparticle 
energies when either the scattering parameter or population imbalance are changed. 
These quantum phase transitions are reflected in the momentum distribution
as well as in thermodynamic properties, however they seem to lie in the non-uniform region
of the phase diagram, but may survive at the center of a harmonic trap~\cite{iskinprl}. 
Furthermore this phase may be observable at finite temperatures in trapped systems~\cite{duan-qpt}, 
or in optical lattices~\cite{iskin-lattice}.

We also analyzed gaussian fluctuations around the saddle point order parameter
both at finite and zero temperatures. 
Near the critical temperature, we derived the Ginzburg-Landau 
equation, and showed that it describes a dilute mixture of composite bosons 
(tightly bound fermions) and excess (unpaired) fermions in the BEC limit. 
At zero temperature, we obtained analytically the dispersion of collective excitations 
in the BCS and BEC limits, and showed numerically the evolution from the BCS to BEC regimes
in the case of zero population imbalance. In addition, we discussed analytically
how phase separation between paired fermions and excess fermions emerges 
analytically at zero temperature in the BEC limit. Lastly, we 
discussed the effects of a harmonic trapping potential, and concluded that
phase separation between paired and unpaired fermions is favored even in
the BEC limit.

\section{Acknowledgements}
\label{sec:acknowledgement}

We thank NSF (DMR-0304380) for support, and S.-K. Yip for e-mail 
correspondence regarding the stability criteria based on the compressibility
matrix.

\appendix

\section{Inverse fluctuation propagator}
\label{sec:app.a}

In this Appendix, we present explicitly the elements of the 
inverse fluctuation propagator $\mathbf{F}^{-1}(q)$.
The diagonal matrix element of $\mathbf{F}^{-1}(q)$ is given by
\begin{eqnarray}
\mathbf{F}_{1,1}^{-1} = \frac{1}{g} &+& 
\sum_{\mathbf{k}} \big\lbrace
v_{\frac{\mathbf{q}}{2} - \mathbf{k}}^2 u_{\frac{\mathbf{q}}{2} + \mathbf{k}}^2 
\frac{n_f(E_{\frac{\mathbf{q}}{2} - \mathbf{k},1}) - n_f(E_{\frac{\mathbf{q}}{2} + \mathbf{k},1})}
{iv_\ell + E_{\frac{\mathbf{q}}{2} - \mathbf{k},1} - E_{\frac{\mathbf{q}}{2} + \mathbf{k},1} } \nonumber \\
&-&
u_{\frac{\mathbf{q}}{2} - \mathbf{k}}^2 v_{\frac{\mathbf{q}}{2} + \mathbf{k}}^2 
\frac{n_f(E_{\frac{\mathbf{q}}{2} - \mathbf{k},2}) - n_f(E_{\frac{\mathbf{q}}{2} + \mathbf{k},2})}
{iv_\ell - E_{\frac{\mathbf{q}}{2} - \mathbf{k},2} + E_{\frac{\mathbf{q}}{2} + \mathbf{k},2} } \nonumber \\
&+&
u_{\frac{\mathbf{q}}{2} + \mathbf{k}}^2 u_{\frac{\mathbf{q}}{2} - \mathbf{k}}^2 
\frac{1 - n_f(E_{\frac{\mathbf{q}}{2} + \mathbf{k},1}) - n_f(E_{\frac{\mathbf{q}}{2} - \mathbf{k},2})}
{iv_\ell - E_{\frac{\mathbf{q}}{2} + \mathbf{k},1} - E_{\frac{\mathbf{q}}{2} - \mathbf{k},2} } \nonumber \\
&-&
v_{\frac{\mathbf{q}}{2} + \mathbf{k}}^2 v_{\frac{\mathbf{q}}{2} - \mathbf{k}}^2 
\frac{1 - n_f(E_{\frac{\mathbf{q}}{2} - \mathbf{k},1}) - n_f(E_{\frac{\mathbf{q}}{2} + \mathbf{k},2})}
{iv_\ell + E_{\frac{\mathbf{q}}{2} - \mathbf{k},1} + E_{\frac{\mathbf{q}}{2} + \mathbf{k},2} } \nonumber \\
&&
\big\rbrace |\Gamma_{\mathbf{k}}|^2
\label{eqn:fupup}
\end{eqnarray}
and the off-diagonal matrix element of $\mathbf{F}^{-1}(q)$ is given by
\begin{eqnarray}
\mathbf{F}_{1,2}^{-1} &=&
\sum_{\mathbf{k}} u_{\frac{\mathbf{q}}{2} + \mathbf{k}} v_{\frac{\mathbf{q}}{2} + \mathbf{k}} 
u_{\frac{\mathbf{q}}{2} - \mathbf{k}} v_{\frac{\mathbf{q}}{2} - \mathbf{k}} \big\lbrace \nonumber \\
&&
\frac{n_f(E_{\frac{\mathbf{q}}{2} - \mathbf{k},1}) - n_f(E_{\frac{\mathbf{q}}{2} + \mathbf{k},1})}
{iv_\ell + E_{\frac{\mathbf{q}}{2} - \mathbf{k},1} - E_{\frac{\mathbf{q}}{2} + \mathbf{k},1} } \nonumber \\
&-&
\frac{n_f(E_{\frac{\mathbf{q}}{2} - \mathbf{k},2}) - n_f(E_{\frac{\mathbf{q}}{2} + \mathbf{k},2})}
{iv_\ell - E_{\frac{\mathbf{q}}{2} - \mathbf{k},2} + E_{\frac{\mathbf{q}}{2} + \mathbf{k},2} } \nonumber \\
&-&
\frac{1 - n_f(E_{\frac{\mathbf{q}}{2} + \mathbf{k},1}) - n_f(E_{\frac{\mathbf{q}}{2} - \mathbf{k},2})}
{iv_\ell - E_{\frac{\mathbf{q}}{2} + \mathbf{k},1} - E_{\frac{\mathbf{q}}{2} - \mathbf{k},2} } \nonumber \\
&+&
\frac{1 - n_f(E_{\frac{\mathbf{q}}{2} - \mathbf{k},1}) - n_f(E_{\frac{\mathbf{q}}{2} + \mathbf{k},2})}
{iv_\ell + E_{\frac{\mathbf{q}}{2} - \mathbf{k},1} + E_{\frac{\mathbf{q}}{2} + \mathbf{k},2} }
\big\rbrace |\Gamma_{\mathbf{k}}|^2
\label{eqn:fupdown}
\end{eqnarray}
where 
$
u_{\mathbf{k}}^2 = (1 + \xi_{\mathbf{k},+}/E_{\mathbf{k},+})/2
$
and
$
v_{\mathbf{k}}^2 = (1 - \xi_{\mathbf{k},+}/E_{\mathbf{k},+})/2,
$
and $n_f(x) = 1/[\exp(\beta x) + 1]$ is the Fermi distribution.

For the s-wave case considered in this manuscript,
a well defined low frequency and long wavelength expansion is possible in two limits: 
(I) at zero temperature ($T = 0$) when population imbalance is zero $P = 0$ such that
the Fermi functions in Eqs. (\ref{eqn:fupup}) and (\ref{eqn:fupdown}) vanish, and 
(II) near the critical temperature ($T \approx T_c$) where $|\Delta_0| \to 0$ such that
$v_{\frac{\mathbf{q}}{2} \pm \mathbf{k}}$ (when $\xi_{\frac{\mathbf{q}}{2} \pm \mathbf{k},+} > 0$) 
and $u_{\frac{\mathbf{q}}{2} \pm \mathbf{k}}$ (when $\xi_{\frac{\mathbf{q}}{2} \pm \mathbf{k},+} < 0$) 
in Eqs. (\ref{eqn:fupup}) and (\ref{eqn:fupdown}) vanish.
Other than these two limits, there is Landau damping which causes the 
collective modes to decay into the two quasiparticle continuum.

\section{Expansion coefficients at $T = T_c$}
\label{sec:app.b}

In this Appendix, we derive the coefficients $a, b, c_{ij}, d$ of the time
dependent Ginzburg-Landau theory described in Eq.~(\ref{eqn:tdgl}). 
We perform a small $\mathbf{q}$ and $iv_\ell \to w + i0^+$ 
expansion of the effective action near the critical temperature ($T \approx T_c$), 
where we assume that the fluctuation field 
$\Lambda(\mathbf{x}, t)$ is a slowly varying function of 
$\mathbf{x}$ and $t$.
The zeroth order coefficient $L^{-1}(0,0)$ is given by
\begin{equation}
a = \frac{1}{g} - \sum_{\mathbf{k}} \frac{X_{\mathbf{k},+}} {2\xi_{\mathbf{k},+}}
|\Gamma_\mathbf{k}|^2
\end{equation}
where 
$
X_{\mathbf{k},\pm} = (X_{\mathbf{k},\uparrow} \pm X_{\mathbf{k},\downarrow})/2
$
and
$
X_{\mathbf{k},\sigma} = \tanh(\beta\xi_{\mathbf{k},\sigma}/2).
$
The second order coefficient $\partial^2 L^{-1}(\mathbf{q},0)/(\partial q_i \partial q_j)$
evaluated at $\mathbf{q} = 0$ is given by 
\begin{eqnarray}
c_{i,j} &=& \sum_{\mathbf{k}} \left[
	\left( \frac{X_{\mathbf{k},\uparrow} Y_{\mathbf{k},\uparrow}}{m_\uparrow^2} 
	+ \frac{X_{\mathbf{k},\downarrow} Y_{\mathbf{k},\downarrow}}{m_\downarrow^2} \right)
	\frac{\beta^2 k_i k_j} {32\xi_{\mathbf{k},+}} \right. \nonumber \\
&+& 
\left( \frac{2k_i k_j	C_-}{m_- \xi_{\mathbf{k},+}} - \delta_{i,j} C_+ \right) \frac{\beta}{16 \xi_{\mathbf{k},+}} \nonumber \\
&+& 
\left. \left(\frac{\delta_{i,j}}{2m_+} - \frac{k_i k_j}{m_-^2\xi_{\mathbf{k},+}}\right) 
	\frac{X_{\mathbf{k},+}} {4\xi_{\mathbf{k},+}^2}
\right] |\Gamma_\mathbf{k}|^2,
\end{eqnarray}
where 
$
C_\pm = (Y_{\mathbf{k},\uparrow}/m_\uparrow \pm Y_{\mathbf{k},\downarrow}/m_\downarrow) / 2,
$
$
$ 
and 
$
Y_{\mathbf{k},\sigma} = {\rm sech}^2(\beta\xi_{\mathbf{k},\sigma}/2).
$
Here, $\delta_{i,j}$ is the Kronecker delta.
The coefficient of the fourth order term is approximated by its value at $\mathbf{q} = \mathbf{0}$, and given by
\begin{eqnarray}
b = \sum_{\mathbf{k}}
\left(
\frac{X_{\mathbf{k},+}}{4\xi_{\mathbf{k},+}^3}
- \frac{\beta Y_{\mathbf{k},+}}{8\xi_{\mathbf{k},+}^2} 
\right) |\Gamma_\mathbf{k}|^4,
\end{eqnarray}
The time-dependent coefficient has real and imaginary parts, and for the $s$-wave case is given by
\begin{eqnarray}
d = \lim_{w \to 0} \sum_{\mathbf{k}}
X_{\mathbf{k},+} 
\left[ \frac{1}{8 \xi_{\mathbf{k},+}^2} + i \pi \frac{\delta(2\xi_{\mathbf{k},+} - w)}{2w}
\right] |\Gamma_\mathbf{k}|^2
\end{eqnarray}
where $\delta (x)$ is the Delta function.

\section{Expansion coefficients at $T = 0$}
\label{sec:app.c}

In this Appendix, we perform a small $\mathbf{q}$ and $iv_\ell \to w + i0^+$ 
expansion of the effective action at zero temperature ($T = 0$).
From the rotated fluctuation matrix ${\mathbf M}^{-1}$ expressed in the amplitude-phase basis, 
we can obtain the expansion coefficients necessary to calculate the collective modes.
We calculate the coefficients only for the case of zero population imbalance $P = 0$, as
extra care is needed when $P \ne 0$ due to Landau damping.
In the long wavelength $(|\mathbf{q}| \to 0)$, and low
frequency $(w \to 0)$ limits the condition
$
\{ w ,\frac{q_i q_j}{2m_+} \} \ll \min \{ 2E_{\mathbf{k},+} \},
$
is used. 

The coefficients necessary to obtain the matrix element 
$\mathbf{M}_{\lambda,\lambda}^{-1} (q)$ are
\begin{equation}
A = \sum_{\mathbf{k}} \frac{|\Delta_0|^2} {2E_{\mathbf{k},+}^3} 
\end{equation}
corresponding to the $(\mathbf{q} = 0, w = 0)$ term,
\begin{eqnarray}
C &=& \sum_{\mathbf{k}} \left[
\xi_{\mathbf{k},+} \frac{E_{\mathbf{k},+}^2 - 3|\Delta_0|^2}{8 m_+ E_{\mathbf{k},+}^5}
- \left( \frac{E_{\mathbf{k},+}^2 - 10|\Delta_0|^2}{m_+^2} \right. \right. \nonumber \\
&+& 
\left. \left. 
\frac{10 |\Delta_0|^4}{m_+^2 E_{\mathbf{k},+}^2} + 
\frac{E_{\mathbf{k},+}^2 - |\Delta_0|^2}{m_-^2} \right)
\frac{k^2}{24 E_{\mathbf{k},+}^5}
\right]
\end{eqnarray}
corresponding to the $|\mathbf{q}|^2$ term, and
\begin{equation}
D = \sum_{\mathbf{k}} \frac{E_{\mathbf{k},+}^2 - |\Delta_0|^2} {8E_{\mathbf{k},+}^5}
\end{equation}
corresponding to the $w^2$ term.

The coefficients necessary to obtain the matrix element
$\mathbf{M}_{\theta,\theta}^{-1} (q)$ are
\begin{equation}
Q = \sum_{\mathbf{k}} \left[
\frac{\xi_{\mathbf{k},+}} {8 m_+ E_{\mathbf{k},+}^3}
- \left( \frac{E_{\mathbf{k},+}^2 - 3|\Delta_0|^2} {m_+^2} + \frac{E_{\mathbf{k},+}^2} {m_-^2} \right) 
\frac{k^2} {24 E_{\mathbf{k},+}^5} 
\right]
\end{equation}
corresponding to the $|\mathbf{q}|^2$ term, and
\begin{equation}
R = \sum_{\mathbf{k}} \frac{1} {8E_{\mathbf{k},+}^3}
\end{equation}
corresponding to the $w^2$ term.

The coefficient necessary to obtain the matrix element 
$\mathbf{M}_{\lambda,\theta}^{-1} (q)$ is
\begin{equation}
B = \sum_{\mathbf{k}} \frac{\xi_{\mathbf{k},+}} {4E_{\mathbf{k},+}^3}
\end{equation}
corresponding to the $w$ term.
These coefficients can be evaluated in the BCS and BEC limits, and
are given in Sec.~\ref{sec:sound-velocity}.

\end{document}